\documentclass{aa}

\usepackage{graphicx}
\usepackage{txfonts}
\newcommand{\gbm}{\textit{Fermi}/GBM}
\newcommand{\xrt}{\textit{Swift}/XRT}

\usepackage{hyperref}
\usepackage{comment}
\usepackage{amsmath}
\usepackage{soul}
\usepackage{booktabs}
\usepackage{url}
\usepackage{multirow}
\usepackage{tcolorbox}
\usepackage{lineno}
\usepackage{mathtools}
\usepackage{caption}
\usepackage{subcaption}

\begin{document} 

   \title{Neutrino search from $\gamma$-ray bursts during the prompt and X-ray afterglow phases using 10 years of IceCube public data}
   \titlerunning{Neutrino search from gamma-ray burst during prompt and X-ray afterglow phases}

   \author{Francesco Lucarelli\inst{1}\fnmsep\thanks{\email{francesco.lucarelli@unige.ch}}\and
           Gor Oganesyan\inst{2,3}\fnmsep\thanks{\email{gor.oganesyan@gssi.it}}\and
           Teresa Montaruli\inst{1}\and
           Marica Branchesi\inst{2,3}\and
           Alessio Mei\inst{2,3}\and
           Samuele Ronchini\inst{2,3}\and
           Francesco Brighenti\inst{2}\and
           Biswajit Banerjee\inst{2,3}
    }
    \institute{
        Département de physique nucléaire et corpusculaire, Université de Genève, CH-1211 Genève, Switzerland\\
        \and
        Gran Sasso Science Institute (GSSI), Via F. Crispi 7, 67100 L'Aquila, Italy\\
        \and
        INFN—Laboratori Nazionali del Gran Sasso, I-67100, L’Aquila (AQ), Italy\\
    }

   \date{}
 
  \abstract{
  Neutrino emission from $\gamma$-ray bursts (GRBs) has been sought for a long time, and stringent limits on the most accredited GRB emission models have been obtained from IceCube. Multi-wavelength GRB observations of the last decades improved our knowledge of the GRB emission parameters, such as the Lorentz factor and the luminosity, which can vary from one GRB to another by several orders of magnitude. Empirical correlations among such parameters have been identified during the prompt phase, with direct implications on GRB models. In this work, we use the PSLab open-access code, developed for IceCube data analyses, to search for individual neutrino emission from the prompt and afterglow phases of selected GRBs, and for stacking emission from the ensemble of such GRBs. For the afterglow phase, we focus in particular on GRBs with X-ray flares and plateaus. While past stacking searches assumed the same GRB fluence at Earth, we present a stacking scheme based on physically motivated GRB weights. Moreover, we conceive a new methodology for the prompt phase that uses the empirical correlations to infer the GRB luminosity and Lorentz factor, when redshift measurements are not available. 
  We do not observe any significant neutrino excess. Hence, we set constraints on the GRB neutrino fluxes and on relevant GRB parameters, including the magnetic field in the jet. Notably, the baryon loading is found to be $<10$ for typical GRB prompts, thus disfavoring a baryonic-dominated origin of the GRB ejecta.
  {}
  {}
  {}
  {}
  {}
  }

   
   \maketitle

    \section{Introduction} \label{sec:intro}

$\gamma$-ray bursts (GRBs) are among the brightest electromagnetic transient events in the Universe. They consist of second- or minute-scale bursts in the soft gamma-ray band, known as prompt phase, followed by a broadband afterglow emission in the radio, optical, and X-rays, lasting a few hours to several days.
The observed photon spectra of the prompt GRB emission are typically fit by two power laws:
\begin{equation}
    \frac{dN_\gamma}{dE_\gamma}\propto
    \begin{dcases}
        \left(\frac{E_\gamma}{E_{\gamma,\rm peak}}\right)^{-\alpha} &\ \ \ E_\gamma<E_{\gamma,\rm peak}\\
        \left(\frac{E_\gamma}{E_{\gamma,\rm peak}}\right)^{-\beta} &\ \ \ E_\gamma>E_{\gamma,\rm peak}\\
    \end{dcases}
\end{equation}
In the double power-law model, the break energy of the spectrum coincides with the peak energy $E_{\gamma,\rm peak}$ of the $E^2_\gamma dN_\gamma/dE_\gamma$ spectrum. Typical spectral indices are $\alpha\simeq0$--1 and $\beta\simeq 2$--3. 

The very fact that the best fit to GRB spectra is a power-law function above $E_{\gamma,\rm peak}$ supports the idea that the GRB spectra are produced by non-thermal phenomena. Nonetheless, twenty years of detailed investigation of the spectra of GRBs did not result in the clear identification of the dominant radiative processes responsible for the prompt emission production. Therefore, a clear understanding of the nature of the GRB jet composition and dissipation processes therein is lacking. However, it is well established, that shocks or re-connection regions produce non-thermal populations of protons and electrons~\citep[for a review, see][]{zhang_2018}. 


Given the non-thermal nature of GRB spectra and their cosmic rate, GRBs are candidate accelerators of ultra-high energy cosmic rays (UHECRs) and are considered promising sources of high-energy neutrinos, as initially proposed in~\cite{Waxman:1997ti}. As a matter of fact, GRBs can provide the sufficient power to justify the spectrum of UHECRs above $10^{17}$~eV, as they are the brightest explosions in the universe with a rate of order of $\sim 1$ per day (as observed by BATSE). Such explosions can release energies as high as $\sim 10^{54}$~erg over time intervals of a few seconds (short GRBs or sGRBs) and around 100~s up to about 1000~s (long GRBs or lGRBs).

The bimodal distribution of the BATSE GRBs hints at two subclasses~\citep{Kouveliotou:1993yx}, which might be generated by two different progenitors, possibly binary mergers for the sGRBs \citep{binary_mergers_sgrb_1989Natur.340..126E} and collapsars for lGRBs \citep{lgrb_doi:10.1146/annurev.astro.43.072103.150558}. This separation/progenitor dichotomy might not be absolute~\citep[see][]{Bromberg_2012ApJ...749..110B}, as the recent detection of a lGRB in coincidence with an identified kilonova light curve proves \citep{2022arXiv220410864R}, followed by evidence of the \textit{Fermi}-LAT observation~\citep{Mei:2022ncd}. 
Recent reviews on GRBs and on models of electromagnetic and neutrino emission can be found for instance in~\cite{Pitik_2021,Kimura_nu_from_GRB:2022zyg}.

The first X-ray afterglow emission was observed by the Beppo-SAX satellite in 1997~\citep{BeppoSAX_Costa:1997obd}. Before the launch of The Neil Gehrels Swift Observatory (\textit{Swift}, hereafter) in 2004, the observational data could be accommodated with an external shock model, in which the afterglow emission was induced by the interactions of the forward and reverse shock of the GRB ejecta with the circum-burst medium~\citep{external_shock_Meszaros_1997}. Nevertheless, the early \textit{Swift} era revealed that a more complex scenario must be considered to explain unexpected features in the X-ray light curve, such as flares and plateaus. 

X-ray flares are rapid rises and falls of the intensity of the X-ray emission observed in nearly 50\% of the GRBs detected by \textit{Swift}~\citep{GRB_Swift_era_doi:10.1146/annurev.astro.46.060407.145147}. They occur tens to hundreds of seconds after the prompt trigger, and last 10~s to $10^3$~s. X-ray flares cannot be produced by external shocks, and they show similarities with the prompt emission~\citep{flares_prompt_link_10.1111/j.1365-2966.2010.16824.x}, which might indicate a common origin powered by a late, long-lasting activity of the central engine~\citep{flares_central_engine2007ApJ...671.1903C}. In this regard, an internal-shock model is proposed at the origin of X-ray flares, in which the production is due to internal dissipation mechanisms~\citep{Internal_shock_model_Fan:2005av}. Given the similarities with the prompt phase and the longer duration, such X-ray flares can lead to a more efficient production of high-energy neutrinos than during the prompt emission, provided that the internal dissipation mechanisms involve a population of non-thermal protons~\citep{Murase_flares_PhysRevLett.97.051101,Kimura_2017ApJ...848L...4K,Kimura_nu_from_GRB:2022zyg}. The high-energy neutrino production is however suppressed if the jet is highly magnetized.

Plateaus are a shallow decay phase of the X-ray emission with time, $\propto t^{-a}$, with a typical temporal slope\footnote{In this work, the term ``plateau'' is used as a synonym of the shallow decay phase. Elsewhere, this term can be used to denote the shallow decay phase only when the temporal slope is close to 0.} $a<0.7$. Plateaus usually occur $10^2$~s to $10^4$~s after the prompt trigger, and they are observed in roughly 60\% of the \textit{Swift} GRBs~\citep{GRB_Swift_era_doi:10.1146/annurev.astro.46.060407.145147}. The plateau can be followed by a normal (slope $1\lesssim a\lesssim2$) or a steep (slope $3\lesssim a\lesssim10$) decay phase in the light curve, with no spectral changes across the break. If followed by a normal decay phase, the plateau can be interpreted within the external shock model~\citep{long_lasting_engine_Zhang_2006}. In this case, the flattening of the X-ray light curve can be attributed to a continuous energy injection supplied by a late-time activity of the central engine~\citep{plateau_central_engine_1998A&A...333L..87D} or to a structured jet~\citep{Oganesyan_2020, Beniamini_2020MNRAS.492.2847B}. The plateau followed by a steep decay phase cannot originate from external shocks~\citep{plateau_steep_decay_Troja:2007ge}. In this regard, many authors believe that the plateau phase is induced by internal dissipation mechanisms powered by a magnetar wind, and that the subsequent steep decay phase is associated with the abrupt collapse of the magnetar to a black hole~\citep{steep_decay_magnetar_10.1111/j.1365-2966.2009.15538.x}.

Neutrino telescopes have conducted many GRB searches. Prominently, IceCube has previously performed searches for spatial and temporal correlation of astrophysical neutrinos and GRBs, finding no significant association. The first exclusion was provided in~\cite{IceCube_GRB_2012:2012qza}, in a search for neutrino emission in time windows up to $\pm 1$~d from GRBs observed from April 2008 to May 2010. In 2015, the analysis of the prompt phase of 506 GRBs in the northern sky from April 2008 to May 2012 yielded constraints on parameters of the single-zone fireball model, and estimated the contribution of GRB neutrinos to $\sim1\%$ of the diffuse astrophysical flux~\citep{IC_GRB_Aartsen_2015}. The analysis was extended in 2017 to include three more years of IceCube data up to May 2015, and prompt GRBs observed in the whole sky~\citep{Aartsen_2017}. An overall catalog of 1172 prompt GRBs was selected for this analysis. More recently, searches for generic extended time windows up to 14~d after the prompt phase and for targeted precursors have been presented in~\cite{GRB_IC_2022}. These searches, that covered more than 2000 GRBs observed from May 2011 to October 2018, place upper limits on the afterglow emission and on the contribution of GRB neutrinos to the diffuse flux, and constrain precursor GRB models. While all the aforementioned analyses adopted an IceCube sample optimized for muon selection, more suitable for point-like source searches (see Sec.~\ref{sec:data}), in~\cite{IC_GRB_Aartsen_2016} IceCube extended the search to include all-flavor high-energy neutrinos detected in three years, from May 2010 to May 2013. A total of 807 GRBs over the entire sky were analysed.

Despite many in-depth investigations of GRB neutrinos, the emission from X-ray plateaus and/or flare afterglows has been constrained by the IceCube collaboration only as a generic afterglow emission. 
In the work described in this paper, we analyze the publicly available IceCube point-source data~\citep{IceCube_data_release:2021xar}, searching for temporal and spatial coincidence of high-energy neutrinos with the prompt phase and with the plateaus and flares of the X-ray afterglow phase to constrain a model inspired from~\cite{Zhang_2013,Kumar2015,Kimura_nu_from_GRB:2022zyg}. We describe this model for clarity in Sec.~\ref{sec:model}. Along with a search from individual GRBs (referred to as ``single-source search'' in the following), we also perform a search for a cumulative neutrino emission (``stacking search''). 

With respect to other IceCube stacking searches, our stacking method adopts physically motivated weights for the contribution of each GRB to the overall emission, thus avoiding the assumption of an equal fluence at Earth. For instance, \cite{Aartsen_2017} provides upper limits on the cumulative neutrino flux from GRBs and constraints on the parameter space of the fireball model, with the underlying assumption that the parameters determining the neutrino emission (e.g. the bulk Lorentz factor $\Gamma$ and luminosity) have the same values for all GRBs. Nonetheless, it is nowadays well established that such parameters can span orders of magnitude for different GRBs.
In this paper, 
motivated by the progress of the last decade in the understanding of the GRB phenomena, and given the increased GRB observations,
we exploit the luminosity--Lorentz factor correlation~\citep{Ghirlanda2012,Lu2012} and the luminosity--peak energy correlation~\citep{Yonetoku_2004ApJ...609..935Y} observed during the prompt phase to determine these parameters on a per-GRB case (see also Sec.~\ref{sec:empirical_correlations}). This approach improves the physical reliability of the results presented in this work compared to past searches.

The paper is structured as follows. In Sec.~\ref{sec:GRBselection}, we introduce the GRB catalogs analysed by this work. Then, we discuss the model and we derive the stacking weights for the different GRB catalogs in Sec.~\ref{sec:model}. We describe the IceCube neutrino sample in Sec.~\ref{sec:data}. We provide details about the statistical method adopted for the single-source and the stacking searches in Sec.~\ref{sec:analysis}. We report the results in Sec.~\ref{sec:results}, and we discuss the related limits and physical implications in Sec.~\ref{sec:discussion}. Final remarks are then summarized in Sec.~\ref{sec:conclusions}.
    \section{Definition of the GRB Catalogs}
\label{sec:GRBselection}



This work analyzes three GRB catalogs. Two catalogs are based on the X-ray afterglow emission, and include GRBs with X-ray plateaus or X-ray flares, hence referred to as ``plateau catalog'' and ``flare catalog'', respectively. A third catalog is based on the prompt phase of selected GRBs, and thus referred to as ``prompt catalog''. We additionally require that the selected GRBs occur during the IceCube uptime provided with the data release~\citep{IceCube_data_release:2021xar}. 
Furthermore, the GRBs with declination $\delta$ above $80^\circ$ in absolute value are excluded, as the technique used to estimate the background of IceCube neutrino events (see Sec.~\ref{sec:analysis}) is not efficient around the equatorial poles.

\begin{figure}[!htb]
    \centering
    \begin{subfigure}{0.4\textwidth}
        \centering

        \includegraphics[width=\textwidth]{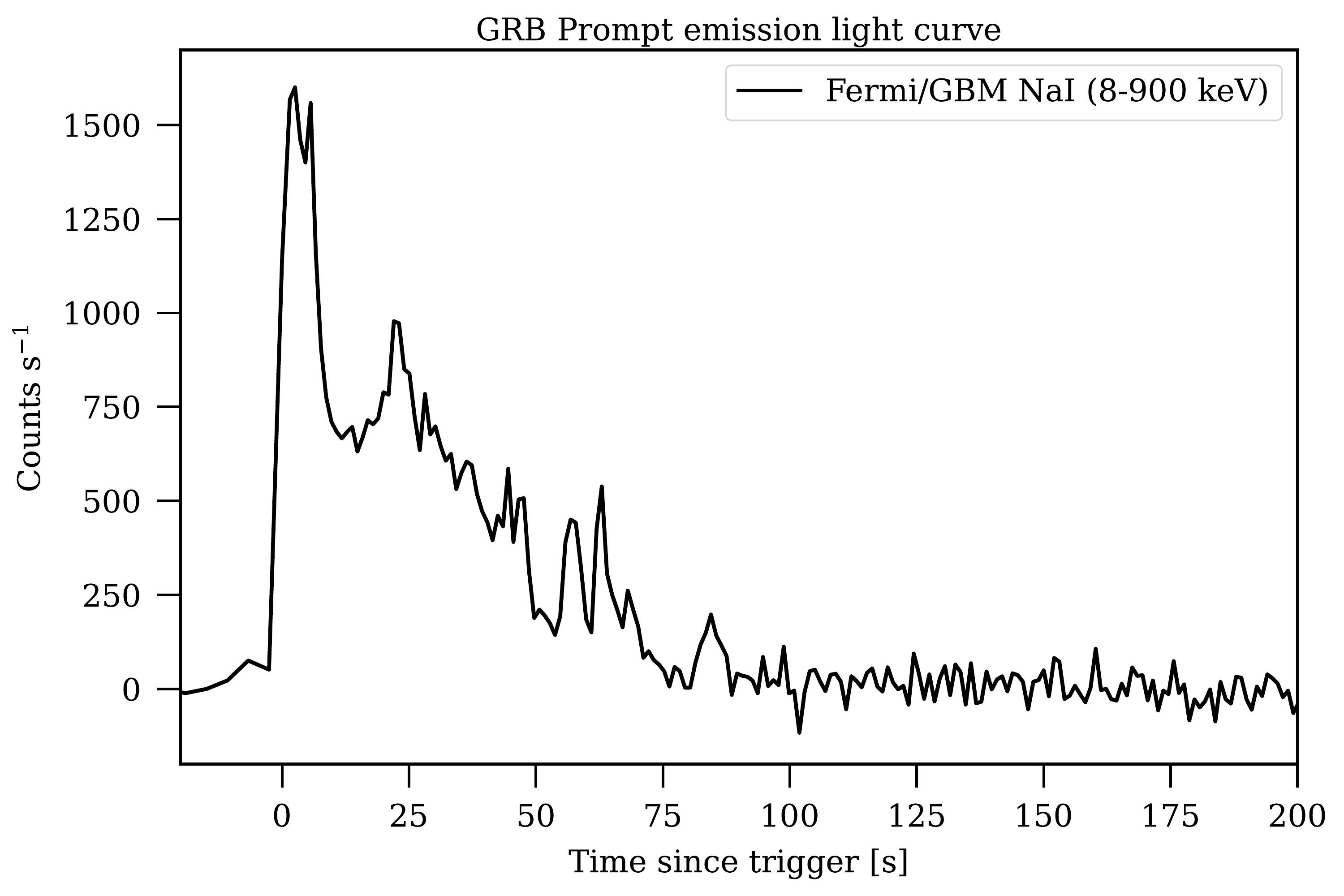}
        \caption{GRB 100727}
        \label{fig:prompt_example}
    \end{subfigure}
    \hfill
    \begin{subfigure}{0.4\textwidth}
        \centering
        \includegraphics[width=\textwidth]{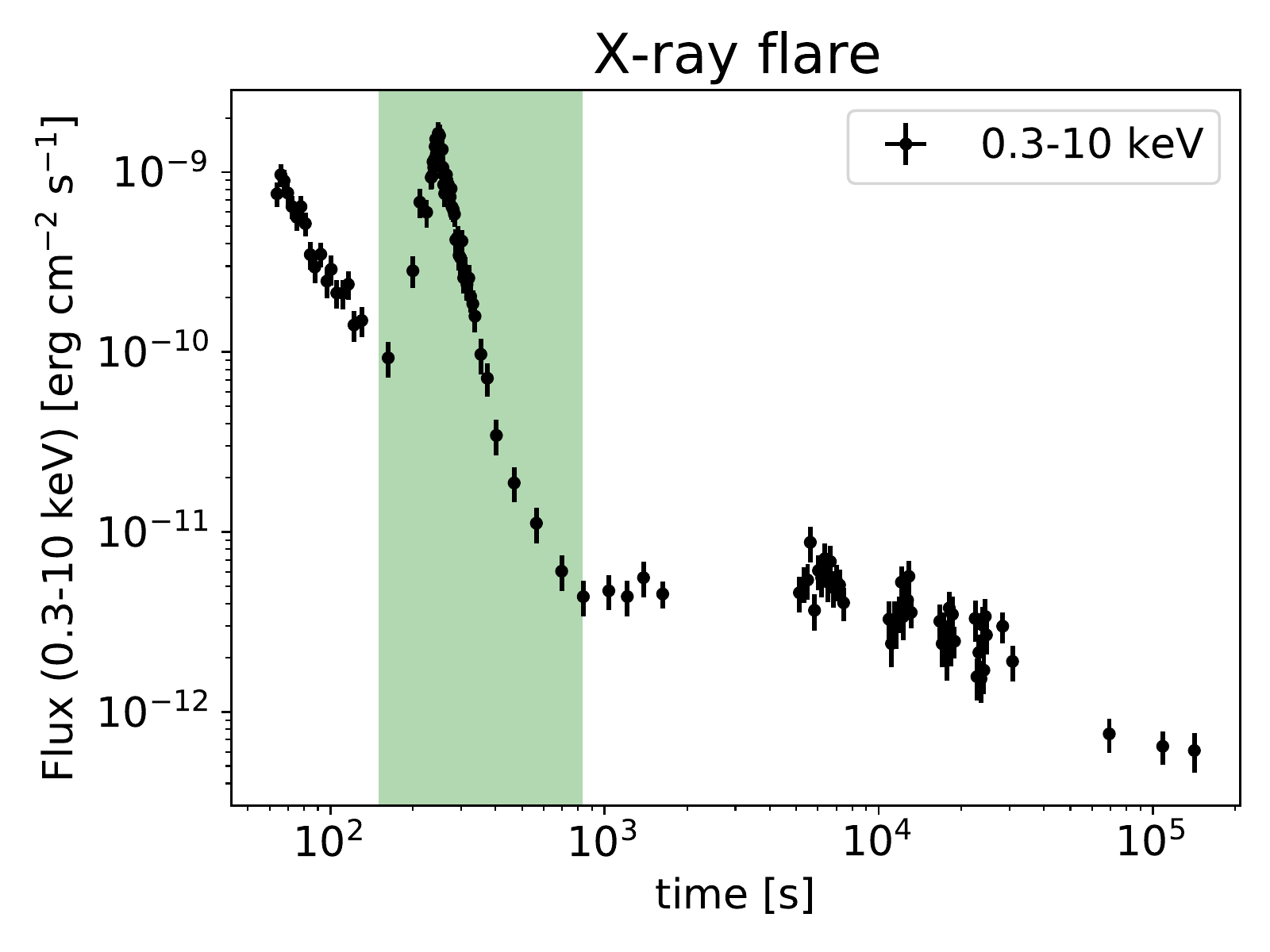}
        \caption{GRB 100727}
        \label{fig:flare_example}
    \end{subfigure}
    \hfill
    \begin{subfigure}{0.4\textwidth}
        \centering
        \includegraphics[width=\textwidth]{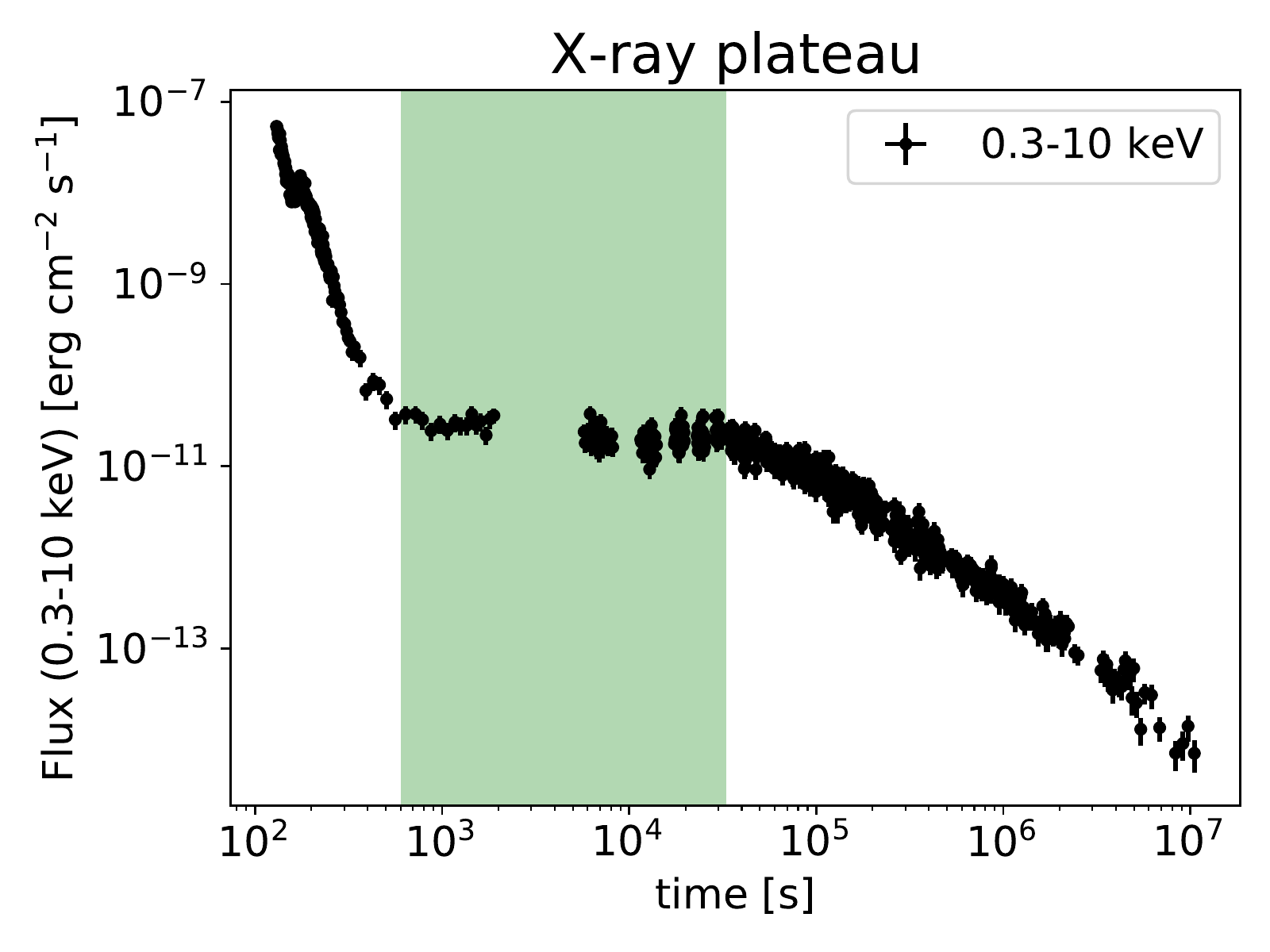}
        \caption{GRB 060729}
        \label{fig:plateau_example}
    \end{subfigure}
    \caption{Examples of a prompt emission light curve  (a) observed by \gbm, an X-ray flare (b) and an X-ray plateau (c) observed by \xrt.}
    \label{fig:flare_plateau_examples}
\end{figure}

The plateau catalog comprises 260 GRBs detected by \xrt, whose X-ray light curve, reconstructed by the automatic analysis of the \xrt\ products~\citep{automatic_analysis_2009MNRAS.397.1177E}, presents a power-law attenuation in time $\propto t^{-a}$, with index $a<0.7$. 
This is the shallowest expected temporal attenuation from synchrotron radiation of external shock~\citep{long_lasting_engine_Zhang_2006}. The flare catalog collects 200 GRBs detected by \xrt, for which the online light curve analysis identifies X-ray flaring activity in the afterglow emission. A total of 85 GRBs have both X-ray plateaus and X-ray flares. They are in common to both catalogs and analyzed considering the corresponding time intervals. Examples of light curves with an X-ray plateau and an X-ray flare are provided in Fig.~\ref{fig:flare_plateau_examples}. The uncertainty on the sky localization of the GRBs that compose these two X-ray-based catalogs benefits from the very good accuracy that characterizes \xrt, and that is generally smaller than the typical angular uncertainty of the IceCube track-like events (see Sec.~\ref{sec:data}).

The 1751 GRBs of the prompt catalog are selected from GRBweb~\citep{GRBweb}, an IceCube project that collects some relevant GRB parameters observed by different telescopes (\xrt~\citep{Swift_catalog_Lien:2016zny}, \gbm~\citep{IPN_catalog_Hurley:2013mf,Fermi_GBM_catalog_vonKienlin:2020xvz}, and others~\citep{Fermi_LAT_catalog_Ajello:2019zki}) into a single, online database. In general, \gbm\ is characterized by a poor angular resolution, that can be as high as $\sim 10^\circ$ and that can exceed the typical resolutions of the track-like neutrino events. In these cases, the GRB is treated as an extended source and its spatial size is taken into account in the analysis method (see Sec.~\ref{sec:analysis} for more details). Nearly $90\%$ of the prompt catalog 
is composed of lGRBs.

The plateau and flare catalogs are analyzed both with the single-source and the stacking searches, described in Sec.~\ref{sec:analysis}, while the prompt catalog is only analyzed with the stacking search. The latter choice is motivated by the fact that IceCube has already performed extensive investigations of GRB neutrinos from the prompt phase~\citep{IceCube_GRB_2012:2012qza,IC_GRB_Aartsen_2015,IC_GRB_Aartsen_2016,Aartsen_2017,GRB_IC_2022}. While our single-source search does not add anything to those searches, we think that a significant difference stems in using the weights defined in Sec.~\ref{sec:model} for the stacking search.
For each catalog, the stacking search is also performed on the sub-samples of GRBs with measured redshift, that will be referred to as ``subcatalogs''. Despite their smaller size, these subcatalogs allow us to reduce the number of assumptions on the GRBs, thus producing less strong, but more reliable results. Tab.~\ref{tab:GRB_catalogs} provides the number of GRBs in each (sub)catalog and in each hemisphere. It also anticipates the stacking weights (derived in Sec.~\ref{sec:model}), defined in terms of the prompt and X-ray isotropic equivalent fluence ($S_\mathrm{iso}$ and $S_\mathrm{iso}^X$) and luminosity ($L_\mathrm{iso}$ and $L_\mathrm{iso}^X$), as well as the peak energy $E_{\gamma,\rm peak}$ and the redshift $z$. Finally, it provides information about the parameters constrained in each catalog, namely the baryon loading factor $\xi_p$, the bulk Lorentz factor $\Gamma$, the variability timescale of the prompt phase $\delta t_\mathrm{obs}$, or the magnetic field $B$ (see also Sec.~\ref{sec:model}).
\begin{table*}[!htbp]
    \centering
    \begin{tabular}{lccccc}
    \toprule
    (sub)catalog & North & South & \multicolumn{2}{c}{weights} & Constrained parameters\\
    \cmidrule(r){4-5}
    & & & $\gamma=1$ & $\gamma=2$ &\\
    \midrule
    \multicolumn{6}{c}{\textbf{Catalogs of GRBs with and without measured redshift}}\\
    \midrule
    Prompt & 959 & 792 & $S_\mathrm{iso}E_{\gamma,\rm peak}^{-1.6}$ & $S_\mathrm{iso}E_{\gamma,\rm peak}^{-1.6}$ & $\xi_p$--$\Gamma$, $\xi_p$--$\delta t_\mathrm{obs}$, $\xi_p$--$B$\\
    X-ray plateau & 141 & 119 & $S_\mathrm{iso}^X$ & $S_\mathrm{iso}^X$ & $\xi_p$--$\Gamma$\\
    X-ray flare & 117 & 83 & $S_\mathrm{iso}^X$ & $S_\mathrm{iso}^X$ & $\xi_p$--$\Gamma$\\
    \midrule
    \multicolumn{6}{c}{\textbf{Subcatalogs of GRBs with measured redshift}}\\
    \midrule
    Prompt & 73 & 51 & $L_\mathrm{iso}S_\mathrm{iso}(1+z)^2$  & $L_\mathrm{iso}S_\mathrm{iso}E_{\gamma,\rm peak}^{-1}$ & $\xi_p$--$\Gamma$, $\xi_p$--$\delta t_\mathrm{obs}$ \\
    X-ray plateau & 63 & 54 & $L_XS_\mathrm{iso}^X(1+z)$ & $L_XS_\mathrm{iso}^X/(1+z)$ & $\xi_p$--$\Gamma$\\
    X-ray flare & 43 & 45 & $L_XS_\mathrm{iso}^X(1+z)$ & $L_XS_\mathrm{iso}^X/(1+z)$ & $\xi_p$--$\Gamma$\\
    \bottomrule
    \bottomrule
    \end{tabular}
    \caption{Description of the GRB catalogs (all-GRB inclusive) and subcatalogs (GRBs with measured redshift) analysed by this work. The number of GRBs is provided separately in each hemisphere, with a separation between the hemispheres at declination $\delta=-5^\circ$. The stacking weights are anticipated here, and provided for a neutrino spectrum $\propto E_\nu^{-\gamma}$; for the derivation of the weights and the definition of the parameters, see Sec.~\ref{sec:model}. The last column of the table shows the 2D space of the GRB parameters constrained by the analysis, as fully detailed in Sec.~\ref{sec:discussion}.}
    \label{tab:GRB_catalogs}
\end{table*}
The catalogs are separated into Northern ($\delta\ge-5^\circ$) and Southern ($\delta<-5^\circ$) hemispheres due to the different sensitivity of IceCube in these two regions of the sky. This is due to the different nature of the background, mostly downgoing muons in the Southern hemisphere, and atmospheric neutrinos in the Northern hemisphere.
    \section{The Model}
\label{sec:model}

\begin{figure*}[!htb]
  \centering
  \includegraphics[width=.7\linewidth]{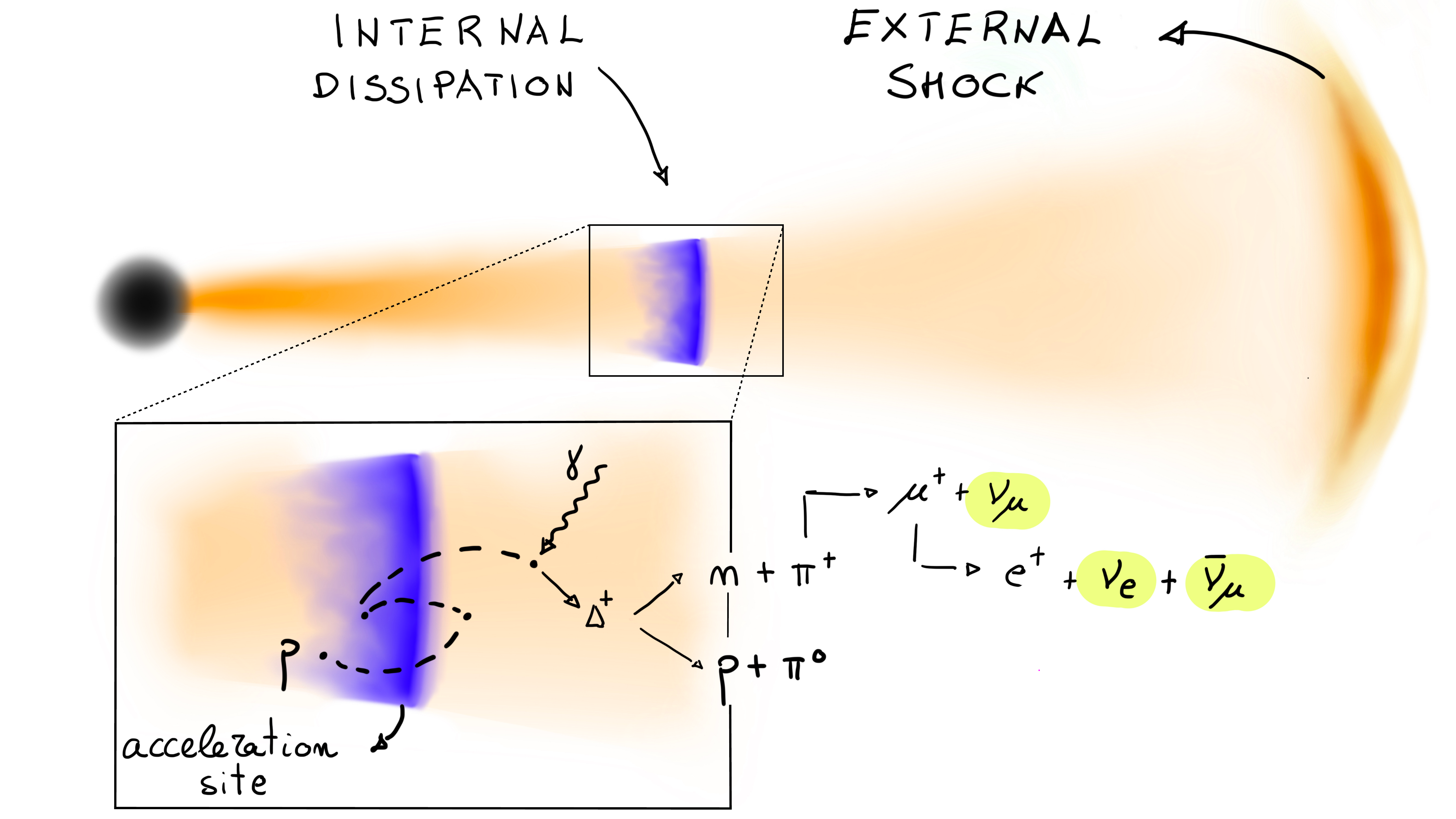}
  \caption{Illustration of the photo-pion processes that lead to the production of high-energy neutrinos inside the dissipation region of GRB jets.}
  \label{fig:photopion_draw}
\end{figure*}

Protons accelerated to high energies can undergo photo-pion production with ambient photons:
\begin{align}
    \label{eq:pgamma_interactions}
    p+\gamma\rightarrow \Delta^+\rightarrow\begin{cases}
        p+\pi^0 \ \ \ \ \ \ \ \mathrm{BR}=2/3~(\mathrm{at~resonance})\\
        n+\pi^+ \ \ \ \ \ \ \ \mathrm{BR}=1/3~(\mathrm{at~resonance})
        \end{cases}
\end{align}
The branching ratios (BR) of the two channels are shown at the resonant energy for the production of an on-shell $\Delta^+$ baryon. Nonetheless, out of resonance the two channels are about equally probable, with $\mathrm{BR}\sim 1/2$ in each. While $\pi^0$'s decay into photon pairs, the decay chain of $\pi^+$'s leads to a copious production of high-energy GRB neutrinos, as also illustrated in Fig.~\ref{fig:photopion_draw}:
\begin{align}
    \label{eq:pi_decay}
    \pi^+ \rightarrow &\mu^+ + \nu_\mu\\
    \label{eq:mu_decay}
    &\mu^+\rightarrow e^+ + \nu_e + \bar{\nu}_\mu
\end{align}
In these photo-pion interactions, each neutrino carries on average 25\% of the parent pion energy. The expected ratio of each neutrino+antineutrino flavor at the source is $(\nu_e:\nu_\mu:\nu_\tau)_\mathrm{S}=(1:2:0)$. Nevertheless, flavor oscillations of TeV--PeV neutrinos over cosmological baselines alter the proportions, and one expects on average the same flux of all flavors at Earth, $(\nu_e:\nu_\mu:\nu_\tau)_\mathrm{E}=(1:1:1)$. Given the chain of reactions above, the emitted neutrino fluence is linked to the proton spectrum $dN_p/dE_p\propto E_p^{-s}$, which depends on the GRB spectral shape.

The proton contribution to the GRB emission is typically expressed in terms of the baryon loading factor $\xi_p=E^\mathrm{tot}_p/E_\mathrm{iso}$, defined as the ratio of the total energy in non-thermal protons $E_p^\mathrm{tot}$ to the isotropic-equivalent bolometric energy released in photons during the prompt phase $E_{\rm iso}$. Typically, $\xi_p\gtrsim 20$ is required for the GRBs to be the sources of the observed UHECR flux \citep[for a review, see][and references therein]{Kimura_nu_from_GRB:2022zyg}. $E_{\rm iso}$ is defined through the observed GRB fluence during the prompt phase $S_\mathrm{iso}$ and the luminosity distance $d_{\rm L}$ at a given redshift $z$:
\begin{equation}
    E_\mathrm{iso}=\frac{4\pi d_L^2S_\mathrm{iso}}{1+z} \, .
\end{equation}
In turn, the GRB fluence $S_{\rm iso}$ is given by the product of the observed GRB flux $F_\mathrm{iso}$ (in units erg~s$^{-1}$~cm$^{-2}$), and the total observation time of the prompt phase $T_{100}$ as defined in GRBweb~\citep{GRBweb}.

In terms of the GRB parameters, the proton spectrum can then be written as~\citep{Kimura_nu_from_GRB:2022zyg}
\begin{equation}\label{eq:protonEnergy}
E_{p}^{2}\frac{dN_p}{dE_{p}} = \xi_{p} f_{p} E_{\rm iso}= \xi_p f_p \frac{4\pi d_L^2S_\mathrm{iso}}{1+z} \, ,
\end{equation}
where $f_p$ is the fraction of energy in protons that produce neutrinos via the photo-meson interactions. It is calculated in Eq.~10 of~\cite{Zhang_2013} for an injection proton spectrum $dN_p/dE_{p} \propto E_{p}^{-s}$ with $s=2$ (that we assume through all this work), and for a range of neutrino energy from 0 to $\infty$. In first approximation, $f_{p}\sim 1/\ln(E_{p, \rm max}/E_{p,\rm min})$, where $E_{p,\rm min}$ and $E_{p,\rm max}$ are respectively the minimum and maximum energy of the non-thermal proton spectrum.

On average, one expects to observe the same flux $\phi_\nu$ of all neutrino flavors at Earth. In the following, we will focus on muon neutrinos and antineutrinos, that constitute the large majority of the events of the IceCube data analyzed by this work (see Sec.~\ref{sec:data}). The muon neutrino+antineutrino fluence $F_{\nu}=\int E^2_{\nu}\phi_{\nu}(E_\nu, t)dt$, at the neutrino energy $E_\nu$, can be linked to the proton spectrum, and ultimately, using Eq.~\ref{eq:protonEnergy}, to the observable GRB parameters~\citep{Kimura_nu_from_GRB:2022zyg}:
\begin{equation}
    \label{eq:nuFluence}
    F_{\nu}=\frac{1}{8}f_{p\gamma}f_\pi^\mathrm{syn}f_\mu^\mathrm{syn}E^2_pN_p(E_p)\frac{1+z}{4\pi d_L^2}=\frac{1}{8} \xi_{p} f_{p} f_{\pi}^{\rm syn} f_{\mu}^{\rm syn}f_{p\gamma} S_{\rm iso} \, .
\end{equation}
The parameters of this equation are described below in this paragraph. The factor 1/8 accounts for the fraction of the pion energy carried by each neutrino ($\sim$1/4) and the branching ratio of non-resonant $p\gamma$ interactions into the $\pi^+$ channel ($\sim$1/2, 
as stated above). $f_{\pi}^{\rm syn}$ and $ f_{\mu}^{\rm syn}$, called pion and muon synchrotron suppression factors, are included in Eq.~\ref{eq:nuFluence} to account for a possible suppression of the high-energy neutrino component due to synchrotron emission of pions and muons that cool down before decaying. The suppression factors are defined as follows:
\begin{equation}\label{eq:f_sync}
f_{i}^{\rm syn} \approx 1-e^{-\frac{t_{i}^{\rm syn}}{t_{i}^{\rm dec}}} \, ,
\end{equation}
where $i$ indicates the particle type, pion or muon. The typical timescale for synchrotron cooling of a particle with energy $E_i$ in a magnetic field $B$ is $t_{i}^{\rm syn}\propto B^{-2}E^{-1}_i$. If this is shorter than the decay timescale $t_{i}^{\rm dec}\propto E_i$, the suppression factors are $f_{i}^{\rm syn}\simeq (E_i^\mathrm{syn}/E_i)^2$, and the neutrino flux can be suppressed at high energy. Normally, this occurs above an energy threshold $E_\pi^\mathrm{syn}\sim10^{18}$~eV for pions and $E_\mu^\mathrm{syn}\sim10^{17}$~eV for muons, assuming typical GRB parameters~\citep{Kimura_nu_from_GRB:2022zyg}. Below these thresholds, the suppression factors are negligible, $f_{i}^{\rm syn}\simeq 1$. Finally, the factor $f_{p\gamma}$ in Eq.~\ref{eq:nuFluence} is the fraction of protons that participate in the photo-pion production. The functional dependency of $f_{p\gamma}$ on the GRB parameters provides an important part of the physical weights of the stacking search, and will be discussed in detail for each (sub)catalog in the following of this section.

The expected GRB neutrino flux is a double broken power law, with a low-energy and a high-energy break point, $\varepsilon_{\nu,\rm br}^{\rm low}$ and $\varepsilon_{\nu,\rm br}^{\rm high}$ respectively~\citep{Waxman:1997ti, Kumar2015}:
\begin{equation}
\label{eq:nuSpectrum}
    \phi_\nu(E_\nu)
    \propto\begin{dcases}
        \left(\frac{E_\nu}{\varepsilon_{\nu,\rm br}^{\rm low}}\right)^{-s+\beta-1} &\ \ \ E_\nu<\varepsilon_{\nu,\rm br}^{\rm low}\\
        \left(\frac{E_\nu}{\varepsilon_{\nu,\rm br}^{\rm low}}\right)^{-s+\alpha-1} &\ \ \ \varepsilon_{\nu,\rm br}^{\rm low}<E_\nu<\varepsilon_{\nu,\rm br}^{\rm high}\\
        \left(\frac{E_\nu}{\sqrt{\varepsilon_{\nu,\rm br}^{\rm low}\varepsilon_{\nu,\rm br}^{\rm high}}}\right)^{-s+\alpha-3} &\ \ \ E_\nu>\varepsilon_{\nu,\rm br}^{\rm high}
    \end{dcases}
\end{equation}

The low-energy neutrino flux break $\varepsilon_{\nu,\rm br}^{\rm low}$ can be calculated from the break of the proton energy, assuming a typical neutrino-to-proton energy ratio of 5\%. This proton break~\citep[see for instance Eq. 29 in][]{Kimura_nu_from_GRB:2022zyg} is due to the photo-pion threshold. Hence, the neutrino low-energy break is:
\begin{equation}
    \varepsilon_{\nu,\rm br}^{\rm low} = 60\left(\frac{\Gamma}{100}\right)\left(\frac{E_{\gamma,\rm peak}}{300~\rm keV}\right)\left(\frac{2}{1+z}\right)^2~\mathrm{TeV} \, .
\end{equation}
Given that IceCube is mostly sensitive in the energy range from few TeV to few PeV, it is very likely to observe GRB neutrinos in the region around $\varepsilon_{\nu,\rm br}^{\rm low}$. The high-energy break $\varepsilon_{\nu,\rm br}^{\rm high}$ is associated to the pion transition from a decay-dominated to a synchrotron-cooling-dominated regime, where the neutrino production is suppressed. Assuming a neutrino-to-pion energy ratio of 25\%, the high-energy break can be estimated as $\varepsilon_{\nu,\rm br}^{\rm high}\simeq0.25E_{\pi}^{\rm syn}\sim10^{17}$~eV. Neutrinos above the high-energy break are unlikely to be observed by IceCube, and therefore they are not considered in this analysis. The neutrino spectral index $\gamma$ in each power-law region is correlated with the proton spectral index $s$ and the photon spectral indices $\alpha$ and $\beta$. Below the low-energy break ($E_\nu<\varepsilon_{\nu,\rm br}^{\rm low}$), neutrinos are produced by the interactions of protons with the high-energy part of the photon spectrum (spectral index $\beta$). In this energy region, the neutrino spectrum scales as $\phi_\nu\propto E_{\nu}^{-\gamma}= E_{\nu}^{-s+\beta-1}$. Above the low-energy break, ($\varepsilon_{\nu,\rm br}^{\rm low}<E_\nu<\varepsilon_{\nu,\rm br}^{\rm high}$), protons can interact with the low-energy photons with spectrum with spectral index $\alpha$, producing a neutrino spectrum $\phi_\nu\propto E_{\nu}^{-\gamma}= E_{\nu}^{-s+\alpha-1}$. Above the high-energy break ($E_\nu>\varepsilon_{\nu,\rm br}^{\rm high}$), the neutrino production is suppressed by synchrotron cooling, and the spectrum is softened by the synchrotron suppression factors, $f_i^\mathrm{syn}\sim E_i^{-2}$. The high-energy part of the neutrino spectrum is thus $\phi_\nu\propto E_{\nu}^{-\gamma}= E_{\nu}^{-s+\alpha-3}$. Assuming a proton spectral index $s=2$ and typical photon parameters $\alpha\simeq 1$ and $\beta\simeq 2$, the corresponding neutrino spectral indices are $\gamma=1$ ($E_\nu<\varepsilon_{\nu,\rm br}^{\rm low}$), $\gamma=2$ ($\varepsilon_{\nu,\rm br}^{\rm low}<E_\nu<\varepsilon_{\nu,\rm br}^{\rm high}$), and $\gamma=4$ ($E_\nu>\varepsilon_{\nu,\rm br}^{\rm high}$), respectively.

\subsection{Empirical Correlations for the Prompt Emission of GRBs}
\label{sec:empirical_correlations}

The stacking weights, that are derived in Sec.~\ref{sec:weights_gamma2} and \ref{sec:weights_gamma1}, depend on GRB parameters, such as the isotropic-equivalent bolometric luminosity $L_\mathrm{iso}$ and the bulk Lorentz factor $\Gamma$. For GRBs with measured redshift, it is trivial to calculate the bolometric luminosity as $L_\mathrm{iso}=4\pi d_L^2 F_\mathrm{iso}$ from the observed GRB flux $F_\mathrm{iso}$ and luminosity distance $d_L$. However, the majority of GRBs do not have measured redshifts. While the past IceCube analyses~\citep[e.g.][]{Aartsen_2017,GRB_IC_2022} adopted the same benchmark values of the relevant parameters for all GRBs, here we propose a novel approach, that exploits empirical correlations observed in lGRBs to estimate the needed quantities for the prompt phase, as first used in~\cite{ANTARES_GRB_analysis_10.1093/mnras/staa3503} to estimate the GRB bulk Lorentz factor. Such an approach benefits from the improvements of the last decade in the understanding of the GRB mechanisms, and it is motivated by the wide sample of GRB observations currently available.

\begin{figure*}[!htb]
    \centering
    \begin{subfigure}{0.45\textwidth}
        \centering
        \includegraphics[width=0.96\linewidth]{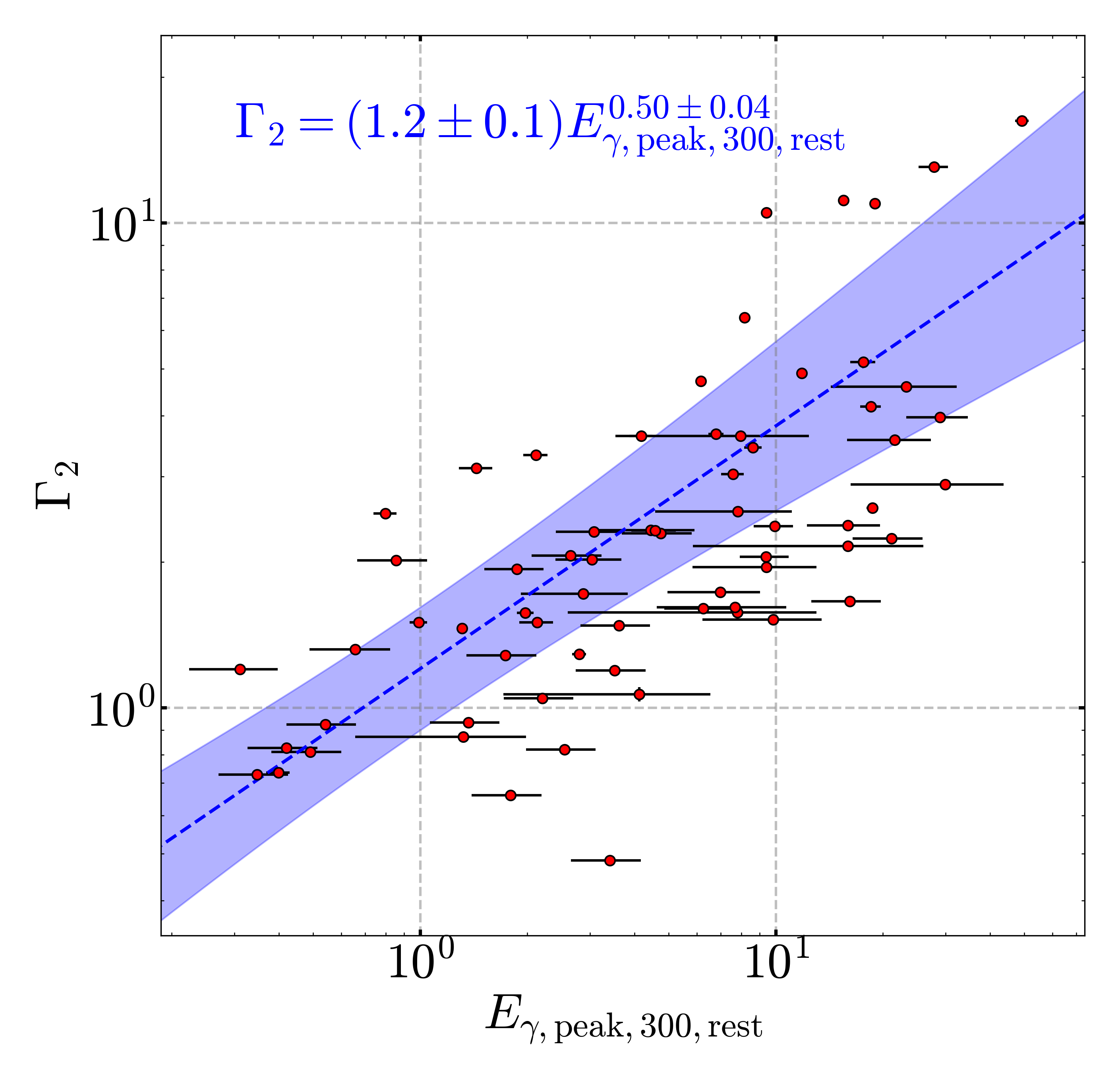}
        \caption{}
    \end{subfigure}
    \begin{subfigure}{0.45\textwidth}
    \centering
        \includegraphics[width=1\linewidth]{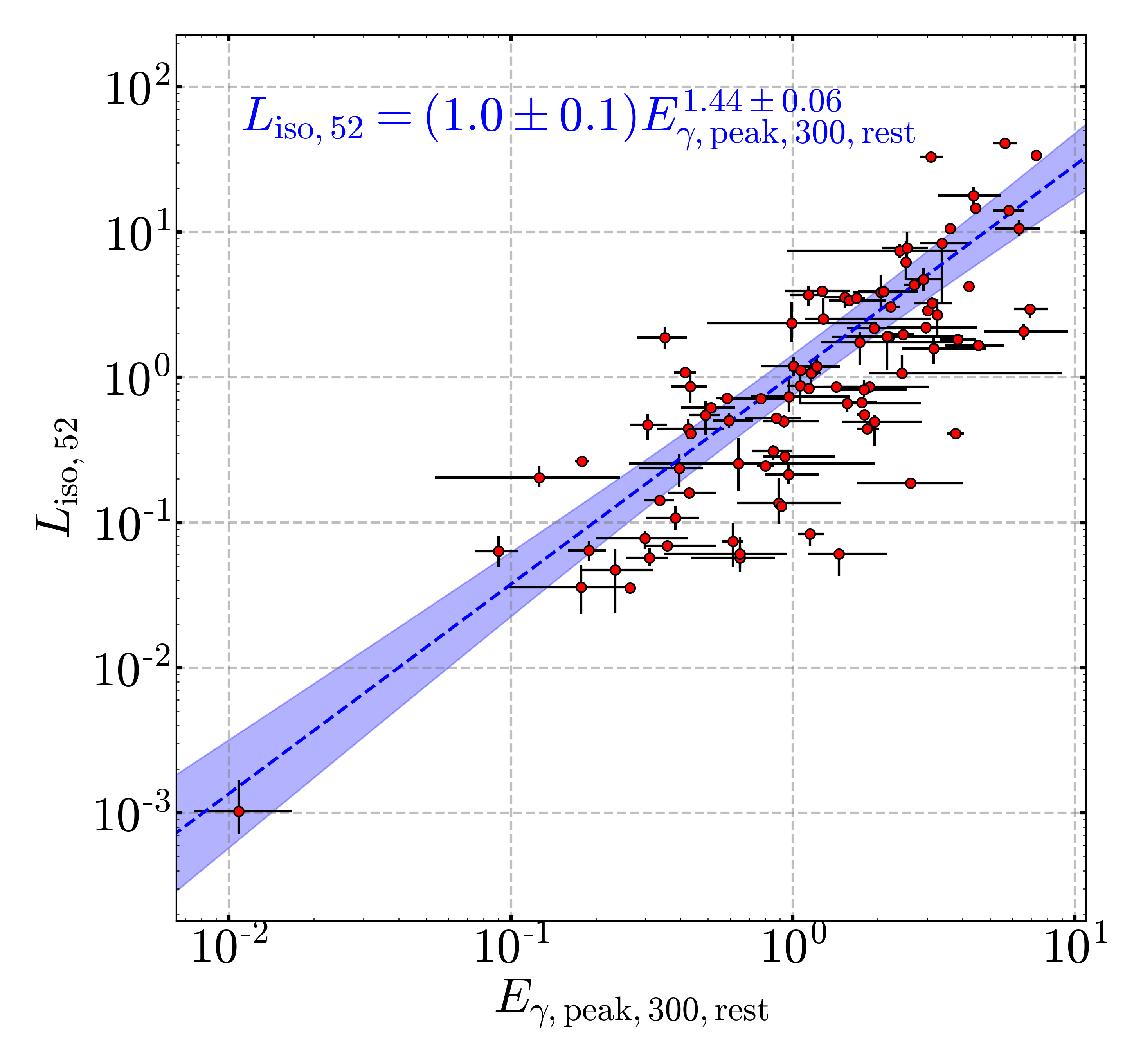}
        \caption{}
    \end{subfigure}
    \caption{Correlation plots of the bulk Lorentz factor $\Gamma_2=\Gamma/10^2$ (a) and luminosity $L_\mathrm{iso,52}=L_\mathrm{iso}/10^{52}$~erg/s (b) with the photon energy peak in the rest frame $E_{\gamma, \rm peak, 300, rest}$ for the prompt emission, useful for the catalog of the GRB prompts. The best exponential fit is shown as a dashed line, with the $3\sigma$ error bands represented by the blue, shaded regions. GRB data are taken from~\cite{ghirlanda2018} (a) and \cite{Yonetoku_10.1093/pasj/62.6.1495} (b).}
    \label{fig:parameters}
\end{figure*}

Two solid correlations have been found between the GRB luminosity and bulk Lorentz factor~\citep{Ghirlanda2012,Lu2012}, and between the GRB luminosity and the photon energy peak in the rest frame~\citep{Yonetoku_2004ApJ...609..935Y}, $E_{\gamma, \rm peak, 300, rest}=(1+z)E_{\gamma, \rm peak, 300}$, where $E_{\gamma, \rm peak, 300}=E_{\gamma, \rm peak}/300$~keV. In Fig.~\ref{fig:parameters} we plot and fit the $\Gamma_2$--$E_{\gamma, \rm peak, 300, rest}$ and the $L_{\mathrm{iso}, 52}$--$E_{\gamma, \rm peak, 300, rest}$ data from the most updated samples in the literature, \cite{ghirlanda2018} and \cite{Yonetoku_10.1093/pasj/62.6.1495} respectively,
where\footnote{In the following, the notation $Q_x$ stands for $Q/10^x$ (in cgs units, for dimensional quantities), with $Q$ being any of the parameters, except for $E_{\gamma,\rm peak,300}$ that corresponds to the normalization at 300~keV} $\Gamma_2 = \Gamma/10^2$ and $L_\mathrm{iso, 52}=L_\mathrm{iso}/10^{52}$~erg/s. Although the original Yonetoku relation in \cite{Yonetoku_2004ApJ...609..935Y} involves the peak luminosity, here we use the (time-averaged) isotropic luminosity, that is more appropriate given that the analysis described in the following of this paper is based on time-averaged GRB properties and on a search for neutrinos from the prompt phase as a whole. For both relationships, the data suggest a linear log-log dependence. In the first case, assuming homogeneous circum-burst medium around the lGRB progenitors (since $\Gamma$ is mainly derived from the peak of the afterglow emission), the best-fit parameters are:
\begin{equation}
    \label{eq:gg_relation}
    \Gamma_{2} = (1.2\pm 0.1)E_{\gamma,\rm peak,300,rest}^{0.50\pm 0.04} \, .
\end{equation}
For the second relationship, the best fit returns:
\begin{equation}
    \label{eq:Yonetoku_relation}
    L_{\rm iso,52} = (1.0\pm0.1) E_{\gamma,\rm peak,300,rest}^{1.44\pm 0.06} \, .
\end{equation}
These relationships are used to estimate the luminosity and Lorentz factor of prompt GRBs for which the redshift is not measured (see below). The relevance of these parameters, namely the peak energy, the luminosity, and the bulk Lorentz factor, to determine the neutrino fluence from individual GRBs was first proposed for GRBs of the BATSE catalogue in the AMANDA era~\citep{Guetta:2003wi}.

\begin{figure}
    \centering
    \includegraphics[width=0.9\linewidth]{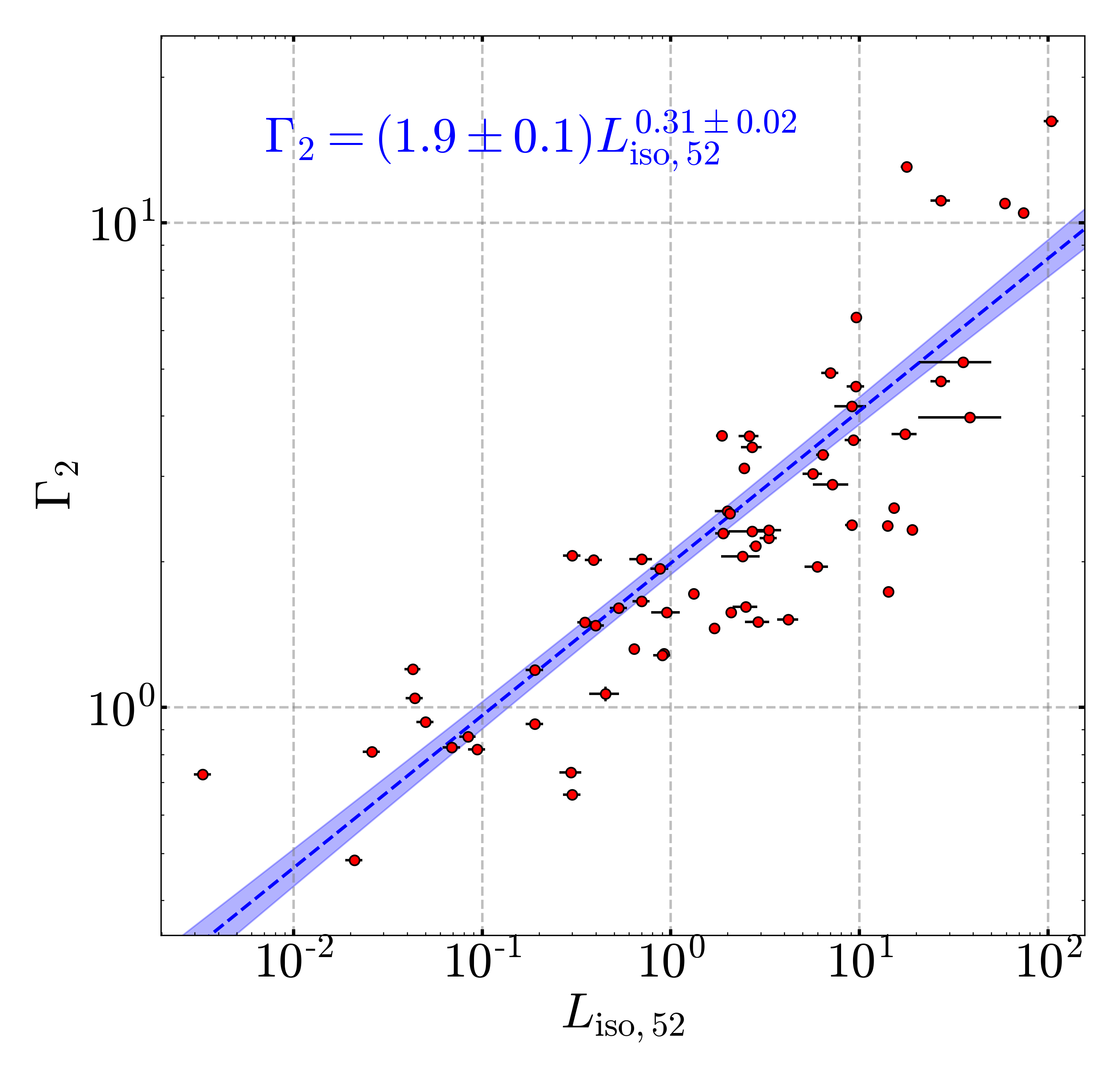}
    \caption{Correlation plot of the bulk Lorentz factor $\Gamma_2=\Gamma/100$ with the isotropic luminosity $L_{\rm iso, 52}=L_{\rm iso}/(10^{52}~\mathrm{erg/s})$ for the prompt emission. The best exponential fit is shown as a dashed line, with the $3\sigma$ error bands represented by the blue, shaded regions. GRB data are taken from~\cite{ghirlanda2018}.}
    \label{fig:Liso_gamma}
\end{figure}

In principle, the bulk Lorentz factor can also be inferred by using the $\Gamma_2$--$L_{\mathrm{iso}, 52}$ relation (fit in Fig.~\ref{fig:Liso_gamma}) together with the $L_{\mathrm{iso}, 52}$--$E_{\gamma, \rm peak, 300, rest}$ relation in Eq.~\ref{eq:Yonetoku_relation}. While leading to the same weighting dependencies derived in Sec.s~\ref{sec:weights_gamma2} and \ref{sec:weights_gamma1}, this approach is more convoluted, and here we prefer to fit and use the direct relation in Eq.~\ref{eq:gg_relation}.

\subsection{Stacking Weights for $\gamma=2$ ($\varepsilon_{\nu,\rm br}^{\rm low}<E_\nu<\varepsilon_{\nu,\rm br}^{\rm high}$)}
\label{sec:weights_gamma2}

In this subsection, we derive the weights for the prompt and X-ray (sub)catalogs in the energy range $\varepsilon_{\nu,\rm br}^{\rm low}<E_\nu<\varepsilon_{\nu,\rm br}^{\rm high}$, where the expected neutrino spectral index is $\gamma=2$. To do so, we estimate the factor $f_{p\gamma}$ in terms of relevant GRB parameters mentioned above and the size of the emission region $R$:
\begin{equation}
\label{eq:f_pgamma}
f_{p\gamma} \approx 2 \chi(\alpha,\beta) \left(\frac{2}{1+z}\right) \frac{L_{\rm iso,52}}{\Gamma_{2}^{2} R_{14}} \frac{1}{E_{\gamma,\rm peak,300}} \, ,
\end{equation}
where $\chi(\alpha,\beta) = \frac{2(2-\alpha)(\beta-2)}{(\beta-\alpha)(1+\beta)}\approx 0.14$ assuming typical $\alpha=1$ and $\beta=2.3$.

In the analysis of the prompt subcatalog, consisting of GRBs with measured redshift, the size of the emission region $R$ can be inferred from the observed time variability $\delta t_\mathrm{obs}$ of the prompt phase~\citep{Piran2004}:
\begin{equation}\label{eq:radius}
R= 2 c \Gamma^{2} \frac{\delta t_{\rm obs}}{1+z} \, .
\end{equation}
Replacing Eq.~\ref{eq:radius} in Eq.~\ref{eq:f_pgamma}, one finds for the prompt emission
\begin{equation}
\label{eq:f_pgamma_prompt}
f_{p\gamma} \approx \frac{2}{3} \chi(\alpha,\beta) \frac{L_{\rm iso,52}}{\Gamma_{2}^{4}}\frac{1}{ \delta t_{\rm obs,0}} \frac{1}{E_{\gamma,\rm peak,300}} \, ,
\end{equation}
where $\delta t_{\rm obs,0}=\delta t_{\rm obs}/1~\rm s$. Combining Eq.~\ref{eq:f_pgamma_prompt} with Eq.~\ref{eq:nuFluence}, and neglecting the synchrotron suppression factors, the expected neutrino fluence is estimated as
\begin{equation}\label{eq:nuFluence_z}
F_\nu  \approx \frac{\chi(\alpha,\beta)}{12\ln(E_{p,\rm max}/E_{p,\rm min})} \left(\frac{L_{\rm iso,52}S_{\rm iso}}{E_{\gamma,\rm peak,300}}\right)\frac{1}{\delta t_{\rm obs,0}}\frac{\xi_{p}}{\Gamma_{2}^{4}} \, .
\end{equation}
The value of $\ln(E_{p,\rm max}/E_{p,\rm min})$ is estimated from simulations. Therefore, the stacking weights for the analysis of the prompt subcatalog are:
\begin{equation}\label{eq:weights_z_prompt}
\omega_{z}^{\gamma} = \frac{L_{\rm iso}S_{\rm iso}}{E_{\gamma,\rm peak}} \, .
\end{equation}
For this subcatalog, we place constraints on $\xi_{p}$ vs $\Gamma$ assuming benchmark values of the time variability $\delta t_{\rm obs}$.

For the analysis of the prompt catalog, where the redshift is not measured for all GRBs and the luminosity and Lorentz factor can be unavailable, we use the correlations in Eq.~\ref{eq:gg_relation} and \ref{eq:Yonetoku_relation} to replace the dependence of $F_\nu$ from $\Gamma_2$ and $L_{\mathrm{iso},52}$ in Eq.~\ref{eq:nuFluence_z} with the observable $E_{\gamma,\rm peak}$. We obtain the following expected neutrino fluence:
\begin{equation}
F_\nu \approx \frac{0.04 \chi(\alpha,\beta)}{\ln(E_{p,\rm max}/E_{p,\rm min})} \left(\frac{S_{\rm iso}}{E_{\gamma,\rm peak,300}^{1.6}}\right) \frac{1}{(1+z)^{0.6}} \frac{\xi_{p}}{\delta t_{\rm obs,0}}\, .
\end{equation}
Therefore, the stacking weights for the analysis of the full prompt catalog (that includes GRBs with no measured redshift) are defined as
\begin{equation}
\omega^{\gamma} = \frac{S_{\rm iso}}{E_{\gamma,\rm peak}^{1.6}}\, .
\end{equation}
In this case, we constrain $\xi_{p}$ vs $\delta t_{\rm obs}$ for fixed values of redshift $z$.

When considering the plateau and flare subcatalogs, the relationship in Eq.~\ref{eq:radius}, used to remove the dependence of $f_{p\gamma}$ from the size $R$ of the emission region during the prompt phase, cannot be used, as the variability timescale for the X-ray flares and plateaus is not defined. In this case, the parameter $R$ cannot be removed from the equations, and the expected neutrino fluence for the X-ray flares and plateaus is
\begin{equation}\label{eq:nuFluence_z_flare}
F_\nu \approx \frac{0.15 \chi(\alpha,\beta)}{\ln(E_{p,\rm max}/E_{p,\rm min})} \left(\frac{L_{X,47}S_{\rm iso}^{X}}{1+z}\right) \frac{1}{E_{\gamma,\rm peak,1}}\frac{1}{R_{14}} \frac{\xi_{p}}{\Gamma_{1}^{2}} \, .
\end{equation}
Here, $L_{X,47}=4\pi d_{L}^{2} F_{X}/(10^{47}~\mathrm{erg/s})$ is the X-ray luminosity normalized to 10$^{47}$~erg/s, $F_{X}$ and $S_{\rm iso}^{X}$ are the observed X-ray flux and fluence respectively, and $E_{\gamma,\rm peak,1}=E_{\gamma,\rm peak}/1$~keV is the energy peak normalized at 1~keV. For the plateau and flare subcatalogs, comprising GRBs with measured redshift, the stacking weights are defined as
\begin{equation}
\omega_{z}^{X} = \frac{L_{X}S_{\rm iso}^{X}}{1+z} \, .
\end{equation}
In this case, the analysis is used to place constraints on $\xi_{p}$ vs $\Gamma$ for benchmark values of $R_{14}$.

For the full plateau and flare catalogs, comprising GRBs with and without measured redshift, the basic weight is used:
\begin{equation}
    \omega^{X} = S_{\rm iso}^{X}\, .
\end{equation}
In this case, no constraints are placed, as too many parameters in Eq.~\ref{eq:nuFluence_z_flare} cannot be estimated in a robust way.

\subsection{Stacking Weights for $\gamma=1$ ($E_\nu<\varepsilon_{\nu,\rm br}^{\rm low}$)}
\label{sec:weights_gamma1}

In this subsection, we derive the weights for the prompt and X-ray (sub)catalogs in the energy range $E_\nu<\varepsilon_{\nu,\rm br}^{\rm low}$, where the expected neutrino spectral index is $\gamma=1$. The neutrino fluence at 1~TeV can be parametrized as
\begin{equation}\label{eq:nuFluence_g1}
F_\nu \approx \frac{2.1 \times 10^{-3} \chi(\alpha,\beta)}{\ln(E_{p,\rm max}/E_{p,\rm min})} L_{\rm iso,52}S_{\rm iso}(1+z)\frac{1}{ R_{14}} \frac{\xi_{p}}{\Gamma_{2}^{4}} \, .
\end{equation}

For the analysis of the prompt subcatalog, the relationship~\ref{eq:radius} is adopted to reduce the dependence of $F_\nu$ in Eq.~\ref{eq:nuFluence_g1} from $R$:
\begin{equation}\label{eq:nuFluence_GRB_g1}
F_\nu \approx \frac{3.5 \times 10^{-4} \chi(\alpha,\beta)}{\ln(E_{p,\rm max}/E_{p,\rm min})} L_{\rm iso,52}S_{\rm iso}(1+z)^{2}
\frac{1}{\delta t_{\rm obs,0}}\frac{\xi_{p}}{\Gamma_{2}^{6}} \, ,
\end{equation}
and the following weights for the GRB prompt emission of GRBs with measured redshifts are defined: 
\begin{equation}\label{eq:weights_z_prompt_g1}
\omega_{z}^{\gamma} = L_{\rm iso}S_{\rm iso}(1+z)^2 \, .
\end{equation}
Constraints are then placed on $\xi_{p}$ vs $\Gamma$ for benchmark values of the variability timescale $\delta t_{\rm obs}$.

When considering the full prompt catalog, the luminosity $L_{\mathrm{iso},52}$ and Lorentz factor $\Gamma_2$ can be replaced with $E_{\gamma,\rm peak}$, using the relationships in Eq.~\ref{eq:gg_relation} and \ref{eq:Yonetoku_relation}:
\begin{equation}\label{eq:nuFluence_GRB_z_g1}
F_\nu \approx \frac{4.1 \times 10^{-6} \chi(\alpha,\beta)}{\ln(E_{p,\rm max}/E_{p,\rm min})}\left( \frac{S_{\rm iso}}{E_{\gamma,\rm peak,300}^{1.6}}\right) (1+z)^{0.4} \frac{\xi_{p}}{\delta t_{\rm obs,0}}
\end{equation}
Therefore, the weights for the analysis of the prompt catalog are 
\begin{equation}
\omega^{\gamma} = \frac{S_{\rm iso}}{E_{\gamma,\rm peak}^{1.6}} \, .
\end{equation}
As for the case with $\gamma=2$, in this case we constrain $\xi_{p}$ vs $\delta t_{\rm obs}$ considering benchmark values of the redshift $z$.

For the analysis of the flare and plateau subcatalogs, we go back to Eq.~\ref{eq:nuFluence_g1}, and we replace the prompt parameters with the corresponding X-ray parameters. We define the following weights:
\begin{equation}
\omega_{z}^{X} = L_{X}S_{\rm iso}^{X}(1+z) \, .
\end{equation}
Constraints are set on $\xi_{p}$ vs $\Gamma$ for some values of $R_{14}$.

For the flare and plateau catalogs, the basic weight is used:
\begin{equation}
\omega^{X} = S_{\rm iso}^{X}. 
\end{equation}
Similarly to the case with $\gamma=2$, no constraints are placed from these two catalogs.

    \section{The IceCube Open-Access Neutrino Data }
\label{sec:data}



The IceCube Neutrino Observatory~\citep{ICNO_Aartsen_2017} is a km$^3$ sized telescope located at the South Pole and designed to detect high-energy ($E>100$~GeV) astrophysical neutrinos. Since 2011, the detector consists of 86 strings embedded in the Antarctic ice at a depth of 1.5--2.5~km, but the data taking period already started in 2006 in partial detector configurations. The strings are equipped with a total of 5,160 digital optical modules, each hosting a 10-inch photomultiplier tube. IceCube detects Cherenkov light from neutrino-induced muon tracks and cascades from neutral and charged current neutrino interactions of tau and electron neutrinos.

The analysis presented in this paper uses the  data collected by IceCube in 10 years of operation, from April 6, 2008 to July 8, 2018, and publicly released as described in~\cite{IceCube_data_release:2021xar}. The data are grouped into five independent sets that reflect the different detector configurations with 40, 59, and 79 strings (from 2008-2010), and the different event selections adopted in the full 86-string configuration for the years 2011 and 2012-2018. Details about the filtering and selection to final level of analyses of the five data sets can be found in Tab.~1 of~\cite{Tessa_analysis_PhysRevLett.124.051103}.

These data are selected for point-source searches, and thus optimized for muon track-like events, that provide the best angular resolution. The sample spans a neutrino energy range between about a hundred GeV to above PeV energies, with median angular resolution below 0.4$^\circ$ above 10~TeV. The roughly 1.1 million events collected in 10 years are mostly dominated by muons from atmospheric neutrino interactions in the Northern Hemisphere (declination $\delta\ge-5^\circ$), and by high-energy atmospheric muons beyond several TeV or large bundles of low-energy muons in the Southern Hemisphere ($\delta<-5^\circ$).
    \section{Analysis Method}
\label{sec:analysis}

For all the searches of this work, we use an unbinned maximum likelihood method similar to the one used for previous point-source IceCube searches~\citep[e.g.][]{Braun:2008bg,Braun_2010,UHECR_nu_correlation,TXS_prior,Tessa_analysis_PhysRevLett.124.051103,Multiflare_IC,TXS_prior,Aartsen_2020,Abbasi_2021}. Such a method is implemented in a public code, known as PSLab,
that has recently been released to the community by the IceCube collaboration. This method is used to search for astrophysical neutrino correlations with individual GRBs (single-source search) and with each (sub)catalog as a whole (stacking search). In the two searches, the tested background hypothesis is the absence of astrophysical neutrinos correlated with individual GRBs for the single-source search, or with the ensemble of GRBs in each (sub)catalog for the stacking search.

\subsection{Single-Source Search}
\label{sec:single_source_analysis}
For a GRB $g$ at direction $\vec{\Omega}_g=(\alpha_g, \delta_g)$, with right ascension $\alpha_g$ and declination $\delta_g$, observed at the instant $t_g$ and lasting $\Delta t_g$, the unbinned likelihood of the single-source search reads as follows:
\begin{align}
    \label{eq:singlesource_lh}
    \begin{split}
        \mathcal{L}^g_j(n_s)=\prod_{i=1}^{N_j}&\left[\frac{n_s}{N_j}\mathcal{S}_{i,j}^g(E_i, \vec{\Omega}_i, \sigma_{i}|\vec{\Omega}_g, t_{g}, \Delta t_{g}, \sigma_{\mathrm{GRB}, g}, \gamma)+\right.\\
        &\left.+\frac{N_j-n_s}{N_j}\mathcal{B}^g_{i,j}(\delta_i, E_i)\right].
    \end{split}
\end{align}
As the 10-yr IceCube data consist of five independent sets, that differ for the detector configuration and event selection criteria, the index $j$ denotes the IceCube data set (from 1 to 5) that contains the time range of the analyzed phase of GRB $g$. The index $i$ runs over the events collected in the IceCube data set $j$, for a total of $N_j$ events. $n_s$ is the amount of signal-like events observed from the tested GRB $g$, and it is the only free likelihood parameter. $\mathcal{S}_{i,j}^g$ and $\mathcal{B}^g_{i,j}$ are the single-source signal and background probability density functions (PDFs) respectively, that depend on the GRB parameters ($\vec{\Omega}_g, t_g, \Delta t_g$) and on the localization error $\sigma_{\mathrm{GRB}, g}$, as well as on the assumed spectral index $\gamma$ of the astrophysical neutrino flux, on the energy proxy $E_i$, on the reconstructed direction $\vec{\Omega}_i=(\alpha_i, \delta_i)$, and on the estimated angular uncertainty $\sigma_i$ of the $i$-th event. 

The signal and background PDFs can be decomposed into the product of a spatial, an energy, and a time PDF. The signal spatial PDF, also called Point-Spread Function (PSF), is modelled as a symmetric bivariate Gaussian distribution, centred at the GRB location and with a variance given by the quadratic sum of the neutrino angular uncertainty and the GRB localization error: $\sigma^2=\sigma_i^2+\sigma_{\mathrm{GRB}, g}^2$. For large standard deviations, $\sigma>10^\circ$, the PSF extends over a non-negligible fraction of the sky, and following the prescription adopted in one of the previous IceCube searches~\citep{Aartsen_2017}, the PSF is replaced by the first-order non-elliptical component of the Kent distribution~\citep{Kent_distribution:10.2307/2984712}:
\begin{equation}
    \frac{\kappa}{4\pi\sinh(\kappa)}e^{\kappa\cos(\Delta\Psi_i^g)}.
\end{equation}
Here, $\Delta\Psi_i^g$ is the angular distance between the reconstructed direction of the event $i$ and the location of the GRB $g$, and $\kappa=1/\sigma^2$ is called concentration parameter. The Kent distribution is the extension of the bivariate Gaussian distribution onto a spherical support, and the two distributions do not differ significantly for large values of the concentration parameter. The signal energy PDF is modeled as a power law with fixed spectral index, $E^{-\gamma}$, as also adopted in both past IceCube searches from GRBs~\citep{Aartsen_2017,GRB_IC_2022}. The analysis is performed by independently assuming $\gamma=1$ or $\gamma=2$, that correspond to the expected shape of the neutrino spectrum below the low-energy and the high-energy break, respectively~\citep{Kimura_nu_from_GRB:2022zyg}. The signal time PDF is constant during the period of the tested GRB, $[t_g, t_g+\Delta t_g]$, and null otherwise. For the analysis of the prompt phase, $\Delta t_g$ is taken as the full prompt duration $T_{100}$, while for the X-ray afterglow it is derived from the automatic analysis of the \xrt\ products~\citep{automatic_analysis_2009MNRAS.397.1177E}.

As the large majority of the IceCube events are of atmospheric origin, the background spatial PDF is determined through a data-driven method, also referred to as ``scramble method'', where the time of the events is randomized within the uptime of the detector, and the right ascension is corrected accordingly, assuming fixed local coordinates (i.e. azimuth and zenith). Due to the peculiar position of the IceCube experiment at the geographical South Pole, the spatial background PDF depends only on the declination, and it is independent of the particular right ascension of the GRBs. Furthermore, since the directional reconstruction efficiency is slightly enhanced if the track events are aligned with the strings of the IceCube detector, an azimuth-dependent correction is included in the background spatial PDF, as described in~\cite{Multiflare_IC}. 
The background energy PDF is also estimated from scrambled data, and it depends on the event declination and energy proxy. The background time PDF is uniform within the time boundaries of each IceCube data set. We neglect the seasonal variations of the atmospheric neutrino and muon background, which is of the order of $\sim 4\%$ 
for neutrino events~\citep{nu_seasonal_variation_Heix:2019jib} and $\sim 8\%$ for downgoing muons~\citep{mu_seasonal_variations_2019arXiv190901406T}.

\subsection{Stacking Search}
The stacking likelihood is based on the same PDFs and parameters as described for the single-source search. It is obtained by extending the single-source likelihood in Eq.~\ref{eq:singlesource_lh} to include all the five IceCube data sets, 
and by replacing the signal PDF with the sum of single-source signal PDFs (for simplicity, the dependence on the neutrino and GRB parameters is omitted):

\begin{equation}
    \label{eq:stacking_lh}
    \mathcal{L}_\mathrm{stack}(n_s)=\prod_{j=1}^{5}\prod_{i=1}^{N_j}\left[\sum_{g\in j}\left(\frac{n_{s,j}^g\mathcal{S}^g_{i,j}}{N_j}+\frac{N_j-n_{s,j}^g}{N_j}\mathcal{B}_{i,j}^g\right)\right].
\end{equation}
The sum runs across all GRBs $g$ occurring in the livetime of the data set $j$. $n_{s,j}^g$ is the expected amount of signal-like events in the data set $j$ originating from GRB $g$, and it is related to the ``cumulative'' signal-like neutrino events $n_s$ through the following equation:
\begin{equation}
\label{eq:individual_GRB_contribution}
n^g_{s,j}=\frac{ w_gR^{\mathrm{IC}}_{j}(\delta_g, \gamma)\Delta T_j}{\sum_j\sum_{g\in j} w_gR_j^\mathrm{IC}(\delta_g,\gamma)\Delta T_j}n_s \, .
\end{equation}
Here, $R^\mathrm{IC}_j(\delta_g,\gamma)$ is the detector acceptance for a neutrino flux $\propto E^{-\gamma}$ related to the data set $j$ at declination $\delta_g$, and $\Delta T_j=\sum_{g\in j}\Delta t_g$ is the overall time extension of all GRBs in the data set $j$. $w_g$ is the theoretical weight on the relative contribution of GRB $g$ to the overall neutrino flux, and it is computed as described in Sec.~\ref{sec:model}. By definition, $n_s=\sum_j\sum_{g\in j} n_{s,j}^g$, i.e. in the stacking search $n_s$ is the amount of signal-like neutrinos from all the stacked GRBs and across the entire analysis period.

\subsection{Test Statistics and \textit{p}-value}
In both the single-source and the stacking searches, the likelihood is maximized with respect to its parameter. The maximum-likelihood parameter is denoted with a hat, $\hat{n}_s$. By setting $n_s=0$, the likelihood describes the background-only hypothesis, and it is thus referred to as background likelihood. The log-likelihood ratio is then used to define the test statistics (TS) of the analysis:
\begin{equation}
    \mathrm{TS}=-2\ln\left(\frac{\mathcal{L}(n_s=0)}{\mathcal{L}(\hat{n}_s)}\right) \, .
\end{equation}
To avoid loss of generality, in the above definition, the likelihood $\mathcal{L}$ is written with no label and it is intended to be the single-source likelihood $\mathcal{L}_j^g$ (Eq.~\ref{eq:singlesource_lh}) or the stacking likelihood $\mathcal{L}_\mathrm{stack}$ (Eq.~\ref{eq:stacking_lh}), depending on the analysis. To quantify the significance of an observed TS, the analyses are repeated for background pseudo-experiments generated with the scramble method, each returning a ``background-like'' TS. The pre-trial $p$-value of an analysis is then estimated as the fraction of background pseudo-experiments that produce a TS value larger than the one observed in the data. When needed, the pre-trial $p$-value is eventually corrected for the so-called ``look-elsewhere effect'' into a post-trial $p$-value that takes into account the number of searches performed. The detail of such a correction are provided in the appropriate Sec.~\ref{sec:results}. In the following, we will refer to cases in which $\mathrm{TS}=0$ as ``under-fluctuations'', and cases in which $\mathrm{TS}>0$ as ``over-fluctuations''.

The major difference of the analysis described in this paper and the previous IceCube analyses of GRB neutrinos~\citep{Aartsen_2017,GRB_IC_2022} lies in the stacking method. In the stacking analysis in~\cite{Aartsen_2017}, the TS is calculated by summing the TS of the individual GRBs from the single-source search. However, the TS distribution is generally declination-dependent, and it might differ for GRBs at different declinations. In~\cite{GRB_IC_2022}, the stacking method coincides with the one adopted for this work, but the individual GRB contributions in Eq.~\ref{eq:individual_GRB_contribution} are only calculated based on the detector acceptance, implicitly assuming an equal GRB fluence at Earth. As a matter of fact, both analyses prefer GRBs that are located in strategic declination bands for IceCube, regardless of the intrinsic properties of the GRBs. In this work, we have proposed a weighted stacking method that, based on physical motivations, estimates the GRB contribution to the expected neutrino fluence by also incorporating relevant GRB observational properties expressed by the weights.

    \section{Results}
\label{sec:results}

The results of the single-source search from the plateau catalog are summarized in Tabs.~\ref{tab:plateau}.
\begin{table*}[!htbp]
    \small
    \centering
    \textbf{Single-Source Plateau Search}\\
    \begin{tabular}{cccccccc}
    \toprule
    GRB name & $\delta$ & $\alpha$ & $t_g$ & $\Delta t_g$ & $\hat{n}_s$ & $p_\mathrm{loc}$ & $F_{90\%}$ \\
     & [ deg ] & [ deg ] & [ MJD ] & [ s ] & & & [ GeV cm$^{-2}$ ] \\
    \midrule
    \multicolumn{8}{c}{\textbf{Spectral index} $\boldsymbol{\gamma=1}$}\\
    \midrule
    GRB 140518A & 42.42 & 227.25 & 56795.390 & $2.9\times10^3$ & 1.0 & $1.1\times 10^{-3}$ (3.1$\sigma$) & $5.5\times 10^{-4}$ \\
    GRB 171120A & 22.46 & 163.79 & 58077.603 & $2.7\times10^4$ & 1.0 & $3.8\times 10^{-3}$ (2.7$\sigma$) & $1.9\times 10^{-4}$ \\
    GRB 120422A & 14.02 & 136.91 & 56039.304 & $1.0\times10^6$ & 0.8 & $2.8\times 10^{-2}$ (1.9$\sigma$) & $9.0\times 10^{-5}$ \\
    GRB 100614A & 49.23 & 263.50 & 55361.954 & $1.4\times 10^5$ & 0.4 & $4.4\times 10^{-2}$ (1.7$\sigma$) & $8.2\times 10^{-4}$ \\
    GRB 180514A & 36.97 & 197.37 & 58252.563 & $1.3\times10^5$ & 0.4 & $5.0\times 10^{-2}$ (1.7$\sigma$) & $4.6\times 10^{-4}$ \\
    GRB 141121A & 22.22 & 122.67 & 56982.431 & $3.1\times10^5$ & 0.5 & $5.0\times 10^{-2}$ (1.7$\sigma$) & $1.8\times 10^{-4}$ \\
    GRB 150213B & 34.19 & 253.45 & 57067.032 & $5.4\times 10^5$ & 0.5 & $7.5\times 10^{-2}$ (1.4$\sigma$) & $4.2\times 10^{-4}$ \\
    \midrule
    \multicolumn{8}{c}{\textbf{Spectral index} $\boldsymbol{\gamma=2}$}\\
    \midrule
    GRB 140518A & 42.42 & 227.25 & 56795.390 & $2.9\times10^3$ & 1.0 & $1.6\times 10^{-3}$ (3.0$\sigma$) & $7.4\times 10^{-2}$ \\
    GRB 171120A & 22.46 & 163.79 & 58077.603 & $2.7\times10^4$ & 1.0 & $3.6\times 10^{-3}$ (2.7$\sigma$) & $7.4\times 10^{-2}$ \\
    GRB 120811C & 62.30 & 199.68 & 56150.652 & $2.3\times10^3$ & 0.1 & $1.9\times 10^{-2}$ (2.1$\sigma$) & $8.2\times 10^{-2}$ \\
    GRB 120422A & 14.02 & 136.91 & 56039.304 & $1.0\times10^6$ & 1.2 & $2.4\times 10^{-2}$ (2.0$\sigma$) & $8.6\times 10^{-2}$ \\
    GRB 180514A & 36.97 & 197.37 & 58252.563 & $1.3\times10^5$ & 0.8 & $4.7\times 10^{-2}$ (1.7$\sigma$) & $7.8\times 10^{-2}$ \\
    GRB 100614A & 49.23 & 263.50 & 55361.954 & $1.4\times 10^5$ & 0.7 & $4.8\times 10^{-2}$ (1.7$\sigma$) & $8.6\times 10^{-2}$ \\
    GRB 150213B & 34.19 & 253.45 & 57067.032 & $5.4\times10^5$ & 1.4 & $4.8\times 10^{-2}$ (1.7$\sigma$) & $9.6\times 10^{-2}$ \\
    GRB 141121A & 22.22 & 122.67 & 56982.431 & $3.1\times10^5$ & 0.7 & $9.7\times 10^{-2}$ (1.3$\sigma$) & $6.9\times 10^{-2}$ \\
    GRB 170306A & -44.75 & 263.07 & 57818.302 & $1.4\times10^5$ & 1.0 & $2.1\times 10^{-2}$ (2.0$\sigma$) & $1.0$ \\
    GRB 130211A & -42.34 & 147.54 & 56334.157 & $6.5\times10^5$ & 0.4 & $7.2\times 10^{-2}$ (1.5$\sigma$) & $9.7\times 10^{-1}$ \\
    \bottomrule
    \bottomrule
    \end{tabular}
    \caption{Results of the single-source analysis of GRBs with X-ray plateau afterglow: GRB name, declination $\delta$, right ascension $\alpha$, starting plateau time $t_g$, plateau duration $\Delta t_g$, maximum-likelihood $\hat{n}_s$, pre-trial $p$-value $p_\mathrm{loc}$, and 90\% confidence-level upper-limit fluence normalization as defined in Eq.~\ref{eq:upper_limit_fluence}. Only GRBs that have a positive TS are shown. A total of 260 GRBs (141 north, 119 south) with plateau are analysed, but only over-fluctuating cases are shown in the table.}
    \label{tab:plateau}
\end{table*}
The search returns GRB 140518A as the most significant GRB, and one signal-like neutrino event identified in temporal coincidence with the X-ray plateau. With a reconstructed muon energy of 3.3~TeV and a directional angular uncertainty of $1.6^\circ$, this event lies $3.3^\circ$ from the location of the source. The corresponding pre-trial $p$-value of GRB 140518A is $p_\mathrm{loc}=1.1\times10^{-3}$ ($3.1\sigma$) for a spectral index $\gamma=1$, and $p_\mathrm{loc}=1.6\times 10^{-3}$ ($3.0\sigma$) for $\gamma=2$. The post-trial $p$-value is calculated by repeating the analysis on many background pseudo-experiments with the corresponding spectral index, and counting the fraction of pseudo-experiments that produces a smaller $p$-value than the one observed in the data. This translates into a post-trial $p$-value of $69\%$ for both spectral indices, thus not significant.

All the GRBs of the X-ray flare catalog are under-fluctuations, except for GRB 120213A when $\gamma=2$ is considered, as reported in Tab.~\ref{tab:flare}. In this case, the data contain one neutrino event in temporal coincidence with the X-ray flare. However, the reconstructed muon energy for this event is only 562~GeV, and the reconstructed track direction is displaced with respect to the source location by $5.8^\circ$, with an angular uncertainty of $2.9^\circ$. This results in a loose association of such neutrino with GRB 120213A, thus explaining the likelihood preference to fit only a fractional event, $\hat{n}_s=0.4$. The pre-trial $p$-value of this GRB is $p_\mathrm{loc}=2.9\times10^{-2}$ ($1.9\sigma$), corresponding to a post-trial $p$-value of $77\%$.
\begin{table*}[!htp]
    \small
    \centering
    \textbf{Single-Source Flare Search}\\
    \begin{tabular}{cccccccc}
    \toprule
     GRB name & $\delta$ & $\alpha$ & $t_g$ & $\Delta t_g$ & $\hat{n}_s$ & $p_\mathrm{loc}$ & $F_{90\%}$ \\
     & [ deg ] & [ deg ] & [ MJD ] & [ s ] & & & [ GeV cm$^{-2}$ ] \\
    \midrule
    \multicolumn{8}{c}{\textbf{Spectral index} $\boldsymbol{\gamma=2}$}\\
    \midrule
    GRB 120213A & 65.41 & 301.01 & 55970.030 & $1.0\times10^4$ & 0.4 & $2.9\times 10^{-2}$ (1.9$\sigma$) & $8.2\times 10^{-2}$ \\
    \bottomrule
    \bottomrule
    \end{tabular}
    \caption{Results of the single-source analysis of GRBs with flare afterglow in X-rays, assuming a spectral index $\gamma=2$: GRB name, declination $\delta$, right ascension $\alpha$, starting flare time $t_g$, flare duration $\Delta t_g$, maximum-likelihood $\hat{n}_s$, pre-trial $p$-value $p_\mathrm{loc}$, and 90\% confidence-level upper-limit fluence normalization as defined in Eq.~\ref{eq:upper_limit_fluence}. Only GRBs with a positive TS are shown. A total of 200 GRBs (117 north, 83 south) with flares are analysed, but only over-fluctuating cases are shown in the table.}
    \label{tab:flare}
\end{table*}

\begin{figure}[!ht]
    \centering
    \begin{subfigure}{0.45\textwidth}
        \centering
        \includegraphics[width=.95\linewidth]{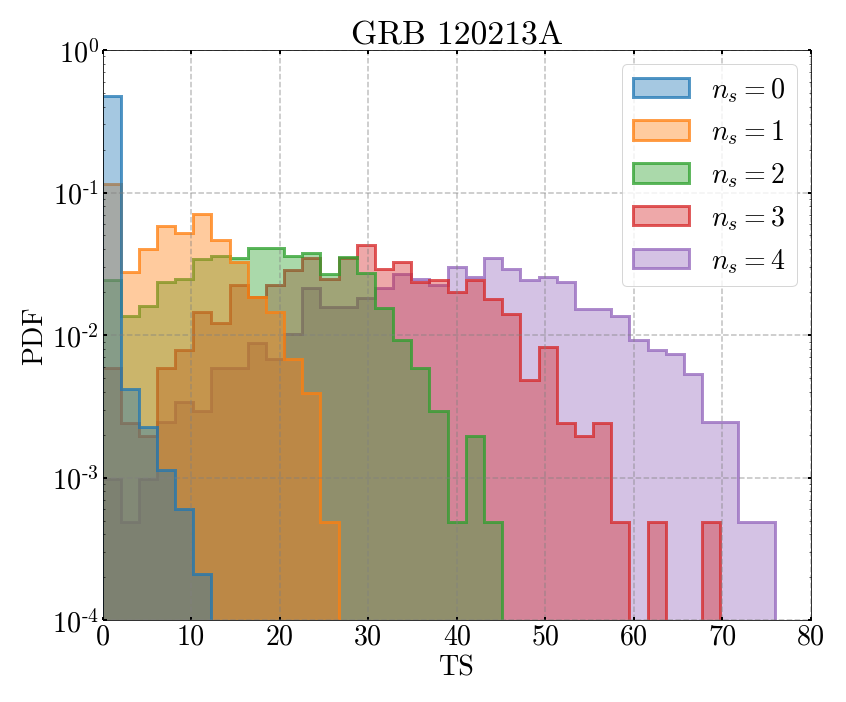}
        \caption{}
    \end{subfigure}
    \begin{subfigure}{0.45\textwidth}
        \centering
        \includegraphics[width=.95\linewidth]{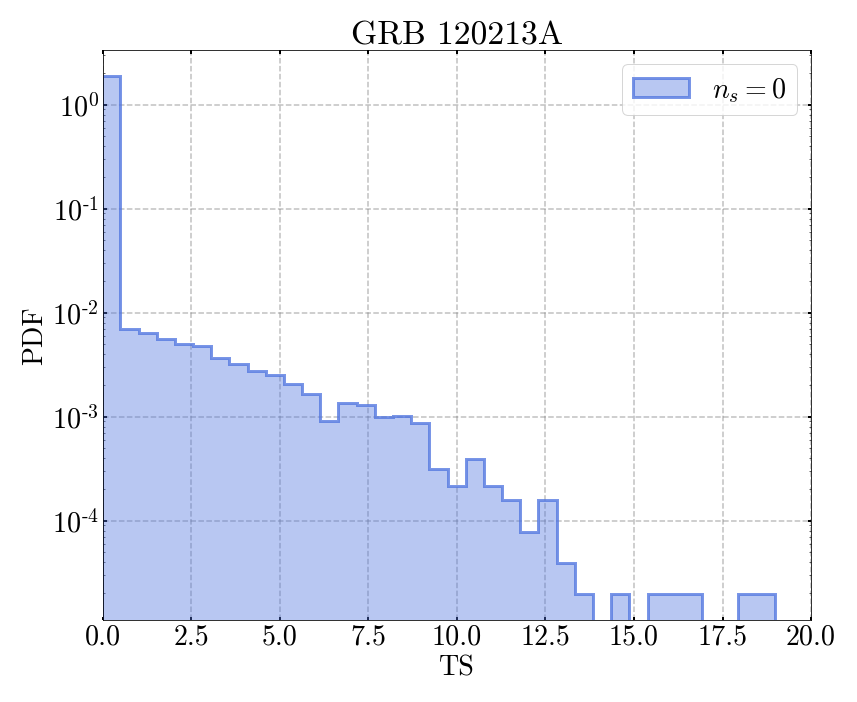}
        \caption{}
    \end{subfigure}
    \caption{(a) Example of TS distributions for the background (injected $n_s = 0$, with 100,000 generated pseudo-experiments) and various signal hypotheses (injected $n_s>1$, with 10,000 pseudo-experiments each) at the location of GRB 120213A, for the single-source search of the flare catalog and assuming spectral index $\gamma=2$. The signal distributions are very different from the background distribution already for $n_s=1$, suggesting that few signal events can lead to a high pre-trial significance. (b) Background TS distribution, the same as in (a), but with different binning and axis ranges. The bin at $\mathrm{TS} = 0$ corresponds to under-fluctuations: for this GRB, about 96\% of the background pseudo-experiments result in under-fluctuations.}
  \label{fig:TS}
\end{figure}

The number of under-fluctuating GRBs ($\sim96\%$ for the plateau catalog and $\sim99.5\%$ for the flare catalog) is much larger than the number of over-fluctuating, and for the flare catalog only one over-fluctuating GRB is identified. This is due to the short duration of the time window in which neutrinos are searched, that reduce the expected amount of events for a fixed event rate. This is particularly true for flares, whose associated search time window ranges between few tens to few hundreds of seconds. In the flare catalog, the expected average under-fluctuation rate under the background hypothesis is $\gtrsim99\%$. However, the very low event rate expected for the flare catalog search makes this analysis almost background-free, suggesting that even the smaller amount of signal can be spotted in the data, potentially carrying a high (pre-trial) significance. This can be easily understood by looking at the background and signal TS distributions of one example GRB from the flare catalog, notably the over-fluctuating GRB 120213A, shown in Fig.~\ref{fig:TS}. These distributions are obtained by running the single-source analysis at the location of GRB 120213A for many background and signal pseudo-experiments: the former are produced with the scramble method described in Sec.~\ref{sec:single_source_analysis}, and the latter by additionally injecting signal-like neutrino events simulated with the PSLab code~\citep{PSLab}.

All the results of the single-source analyses are compatible with the background hypothesis. As such, 90\% confidence-level (CL) upper limits on the neutrino fluence from individual non-underfluctuating sources are placed. The upper-limit neutrino fluence for the single-source searches is defined as follows:
\begin{equation}
    \label{eq:upper_limit_fluence}
    F_\nu(E_\nu)=E^2\frac{\mathop{dN_\nu}}{\mathop{dE_\nu}\mathop{dA}}=F_{90\%}\left(\frac{E_\nu}{\mathrm{TeV}}\right)^{2-\gamma}.
\end{equation}
The factor $F_{90\%}$, corresponding to the normalization of the upper-limit fluence at the benchmark value of 1~TeV, is reported in the last column of Tab.s~\ref{tab:plateau} and \ref{tab:flare} for the related sources.

The results of the stacking searches are summarized in Tab.~\ref{tab:stacking}.
\begin{table*}[!htbp]
    \small
    \centering
    \textbf{Stacking Search}\\
    \begin{tabular}{cccccccc}
    \toprule
     \multirow{2}{*}{Catalog} & \multirow{2}{*}{Hemisphere} & \multicolumn{3}{c}{$\gamma=1$} & \multicolumn{3}{c}{$\gamma=2$} \\
     \cmidrule(r){3-5} \cmidrule(l){6-8}
     & & $\hat{n}_s$ & $p_\mathrm{loc}$ & $\phi^\mathrm{Stack}_{90\%}$ & $\hat{n}_s$ & $p_\mathrm{loc}$ & $\phi^\mathrm{Stack}_{90\%}$\\[3pt]
      & & & & [ GeV cm$^{-2}$ s$^{-1}$ sr$^{-1}$ ] & & & [ GeV cm$^{-2}$ s$^{-1}$ sr$^{-1}$ ]\\
    \midrule
    \multicolumn{8}{c}{\textbf{Catalogs of GRBs with and without measured redshift}}\\
    \midrule
    \multirow{2}{*}{Prompt} & North & -- & -- & $2.2\times10^{-14}$ & 0.9 & $3.7\times10^{-2}$ & $3.5\times10^{-11}$\\
     & South & -- & -- & $8.0\times10^{-15}$ & -- & -- & $1.2\times10^{-10}$ \\
    \midrule
    \multirow{2}{*}{Plateau} & North & -- & -- & $8.6\times10^{-14}$ & -- & -- & $5.1\times10^{-11}$ \\
     & South & -- & -- & $2.0\times10^{-14}$ & -- & -- & $4.1\times10^{-10}$ \\
     \midrule
     \multirow{2}{*}{Flare} & North & -- & -- & $7.0\times10^{-15}$& -- & -- & $4.1\times10^{-11}$ \\
     & South & -- & -- & $1.7\times 10^{-14}$ & -- & -- & $3.5\times10^{-10}$\\
    \midrule
    \multicolumn{8}{c}{\textbf{Subcatalogs of GRBs with measured redshift}}\\
    \midrule
    \multirow{2}{*}{Prompt} & North & -- & -- & $1.1\times10^{-14}$ & -- & -- & $2.5\times 10^{-11}$ \\
     & South & -- & -- & $1.5\times10^{-14}$ & -- & -- & $1.9\times10^{-10}$ \\
    \midrule
    \multirow{2}{*}{Plateau} & North & -- & -- & $1.0\times10^{-13}$ & -- & -- & $3.8\times10^{-11}$ \\
     & South & -- & -- & $2.3\times10^{-14}$ & -- & -- & $5.1\times10^{-10}$ \\
     \midrule
     \multirow{2}{*}{Flare} & North & -- & -- & $2.5\times10^{-14}$ & -- & -- & $3.4\times10^{-11}$ \\
     & South & -- & -- & $1.2\times10^{-14}$ & -- & -- & $2.2\times10^{-10}$ \\
    \bottomrule
    \bottomrule
    \end{tabular}
    \caption{Results of the stacking analysis of the three GRB catalogs (prompt, plateau, flare) in each hemisphere, with and without requirements on the GRB redshift. For each value of the tested spectral index, $\gamma=1$ and $\gamma=2$, the best-fit $\hat{n}_s$, the pre-trial $p$-value $p_\mathrm{loc}$, and the upper limits on the stacking fluence at 90\% confidence level $\phi_{90\%}^\mathrm{Stack}$ as defined in Eq.~\ref{eq:upper_limit_flux} are reported. Underfluctuations are shown with hyphens.}
    \label{tab:stacking}
\end{table*}
These searches produce underfluctuations in all the (sub)catalogs, except for the analysis of the prompt catalog in the Northern Hemisphere with spectral index $\gamma=2$, with no requirements on the measured redshift. In this case, the stacking likelihood fits $\hat{n}_s\simeq 1$ event, resulting in a pre-trial $p$-value of $p_\mathrm{loc}=3.5\times10^{-2}$ and a post-trial $p$-value of 13\%. The post-trial correction of the stacking search is due to performing several searches of this kind, on different hemispheres, on different catalogs, with different spectral indices ($\gamma=1$ and $\gamma=2$), and with different requirements on the measured GRB redshift.

\begin{figure*}[!htb]
  \centering
  \includegraphics[width=.49\linewidth]{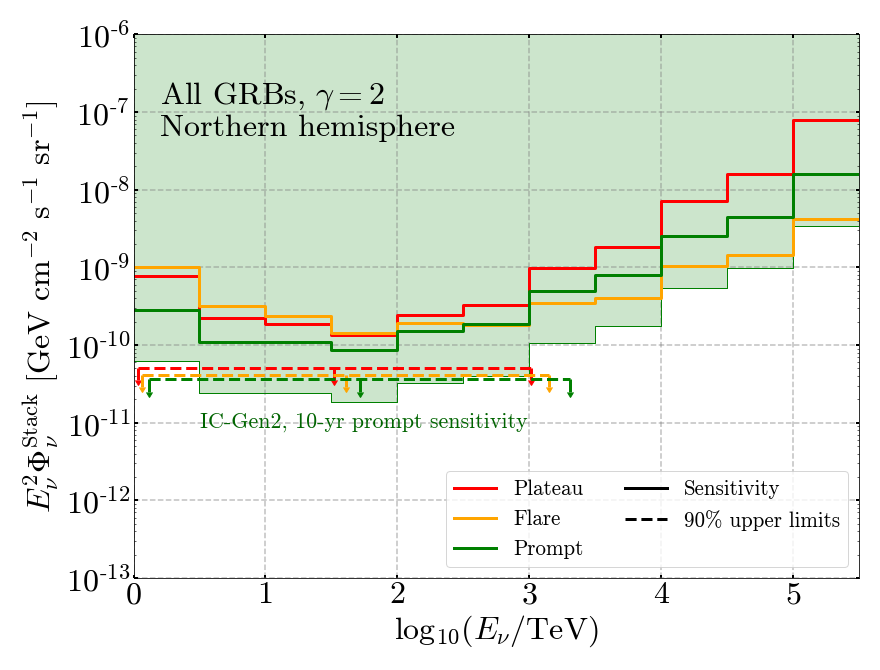}%
  \includegraphics[width=.49\linewidth]{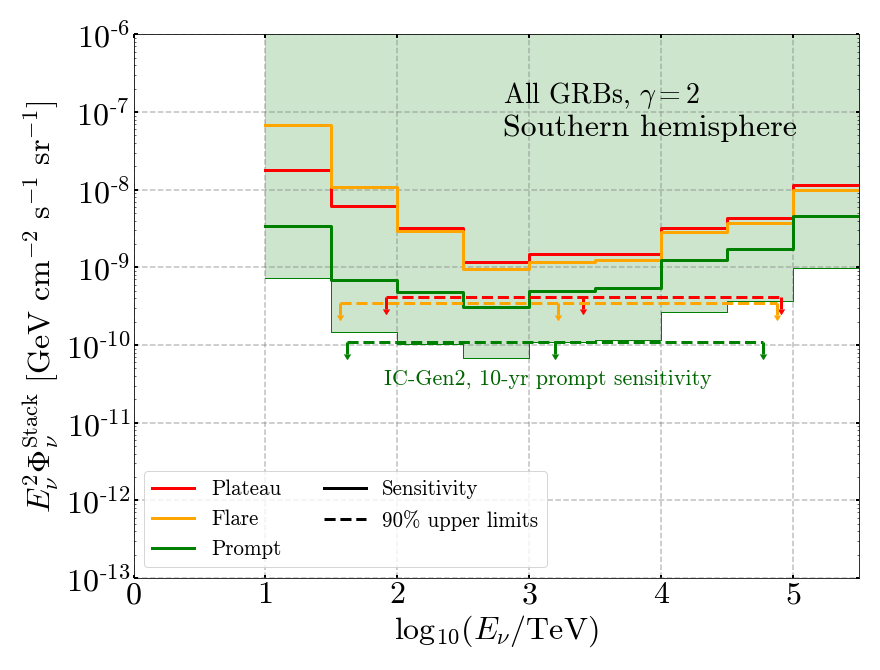}
  \includegraphics[width=.49\linewidth]{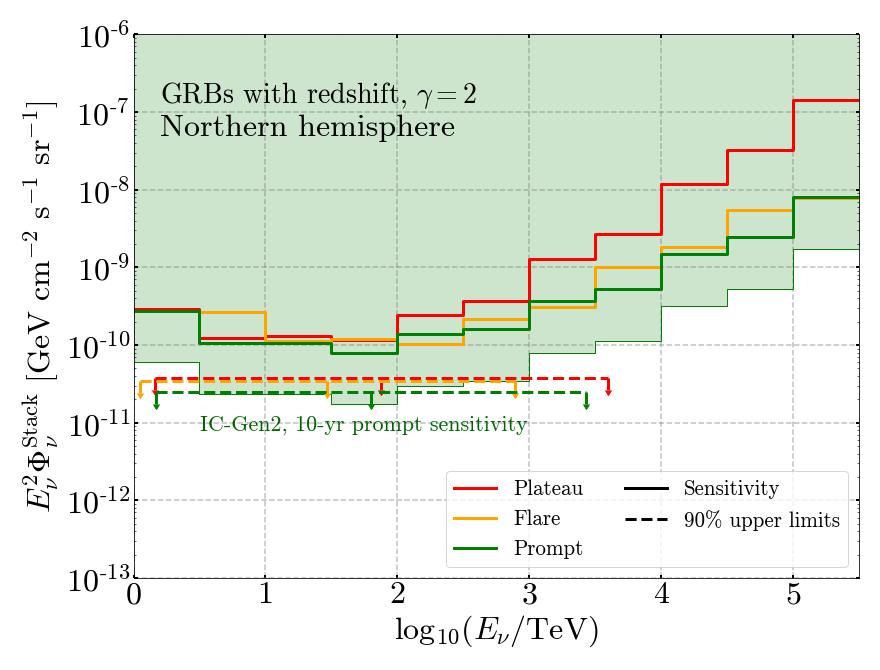}%
  \includegraphics[width=.49\linewidth]{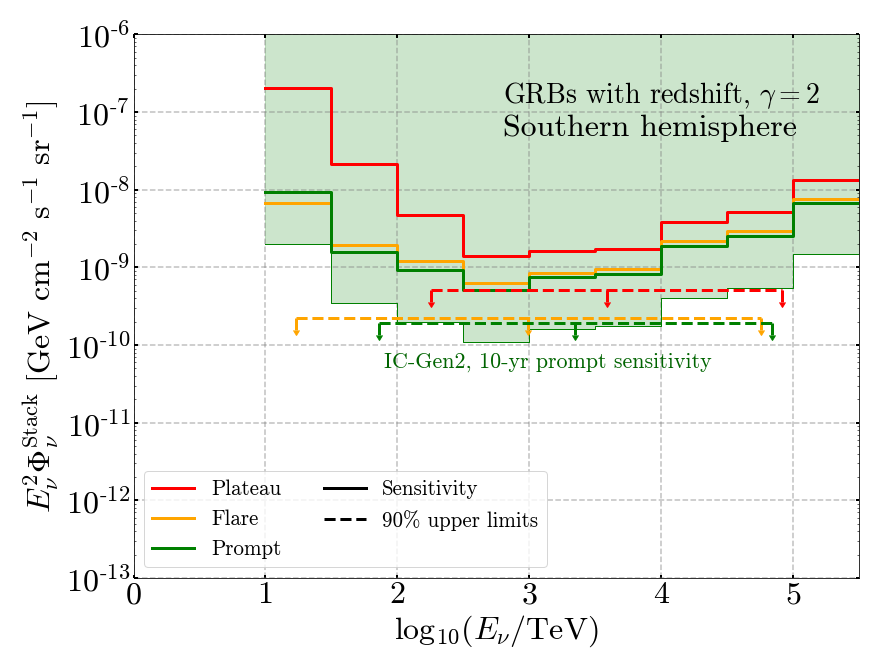}
  \caption{Differential sensitivity (solid lines) and 90\% CL upper limits (dashed lines) for the stacking analysis of the plateau (red), flare (yellow), and prompt (green) catalog, assuming $\gamma=2$. Curves are shown for the Northern (left) and Stacking (right) hemisphere, and for catalogs comprising all the GRBs (first row) or the subcatalogs of GRBs with measured redshift (second row). As a comparison, the shaded green region shows the 10-yr differential sensitivity calculated for IceCube-Gen2~\citep{IC_Gen2_Clark_2021} as explained in the text.}
  \label{fig:sens_upLims_gamma2}
\end{figure*}

As no significant astrophysical correlation is observed by the stacking searches, these analyses are used to place 90\% CL upper limits on the cumulative neutrino flux from each (sub)catalog. The upper-limit neutrino flux for the stacking searches is defined as follows:
\begin{align}
    \label{eq:upper_limit_flux}
    \begin{split}
        E_\nu^2\phi^\mathrm{Stack}_\nu(E_\nu)&=E^2\frac{\mathop{dN^\mathrm{Stack}}}{\mathop{dE}\mathop{dA}\mathop{dt}\mathop{d\Omega}}=\\
        &=F^\mathrm{Stack}_{90\%}\left(\frac{E_\nu}{\mathrm{TeV}}\right)^{2-\gamma}\frac{1}{\Delta T}\frac{1}{\Delta\Omega}=\phi^\mathrm{Stack}_{90\%}\left(\frac{E_\nu}{\mathrm{TeV}}\right)^{2-\gamma},
    \end{split}
\end{align}
where $\Delta T\simeq10$~yr is the full livetime of the data used for the analysis, and $\Delta\Omega\simeq 2\pi$ is the solid angle of each hemisphere. $F^\mathrm{Stack}_{90\%}$ and $\phi^\mathrm{Stack}_{90\%}$ are the normalizations at 1~TeV on the 90\% CL upper limit of the neutrino stacking fluence and the neutrino stacking flux, respectively. The 90\% CL upper-limit flux normalization for each (sub)catalog is reported in Tab.~\ref{tab:stacking}. For a better visualization, such upper limits are also displayed in Fig.~\ref{fig:sens_upLims_gamma2}, assuming the same neutrino spectrum $\propto E_{\nu}^{-2}$ for all the GRBs. This figure additionally shows the energy-dependent stacking sensitivity of this analysis, and a comparison of the estimated IceCube-Gen2 sensitivity for the prompt (sub)catalog, assuming an $E_\nu^{-2}$ neutrino spectrum. The IceCube-Gen2~\citep{IC_Gen2_Clark_2021} sensitivity is calculated by considering a 10 times larger detector volume, and by scaling the current effective area (and hence the sensitivity) by a factor $10^{2/3}$, close to the designed factor $\sim5$~\citep{IceCube_gen2:2014gqr}. The flux in Eq.~\ref{eq:upper_limit_flux} follows a similar definition as for the IceCube search in~\cite{GRB_IC_2022}, but it differs from the previous IceCube search in~\cite{Aartsen_2017}.
    \section{Discussion}
\label{sec:discussion}

The upper limits derived in Sec.~\ref{sec:results} from the non-observation of significant events in the various stacking searches are used to constrain relevant quantities of the model discussed in Sec.~\ref{sec:model}. 
Our assumption for the prompt emission is based on the empirical correlations between the GRB luminosities and the peak energies, and between the GRB bulk Lorentz factors and the peak energies, as detailed in~\cite{Yonetoku_10.1093/pasj/62.6.1495} and \cite{Ghirlanda2012}, respectively. Unlike previous GRB analyses performed by IceCube, that assumed all GRBs to have the same bulk Lorentz factor, fluence, or luminosity, we take into account the large range of values for these parameters (e.g. the luminosity spans five orders of magnitude). We assume the aforementioned correlations to estimate realistic values for these parameters of each GRB.

\begin{figure}[!htb]
  \centering
  \includegraphics[width=1.\linewidth]{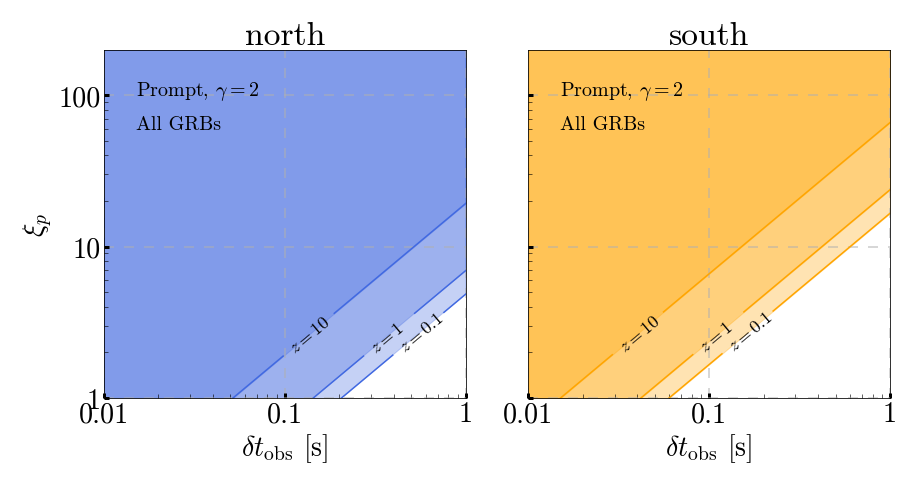}
  \includegraphics[width=1.\linewidth]{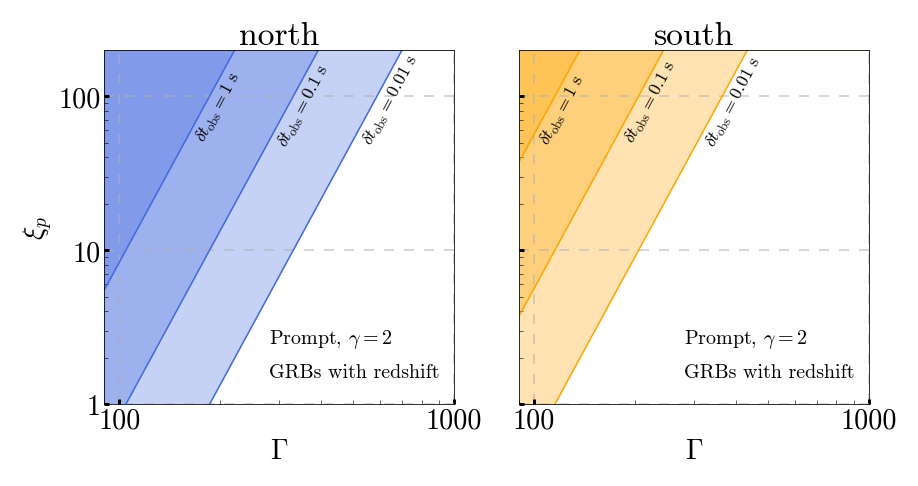}
  \caption{$90\%$ CL upper limits for the baryon loading factor $\xi_p$ in the Northern (left) and Southern (right) Sky. These constraints are obtained with the stacking analysis of the prompt catalog (upper plots) and subcatalogs of GRBs with available redshift (bottom plots), assuming a neutrino spectral index $\gamma=2$. Excluded values are shown as shaded regions for different hypotheses of the redshift $z$ (top plots) and of the timescale variability $\delta t_\mathrm{obs}$ (bottom plots).}
  \label{fig:gamma2_prompt_constraint}
\end{figure}

\begin{figure}[!htb]
  \centering
  \includegraphics[width=1.\linewidth]{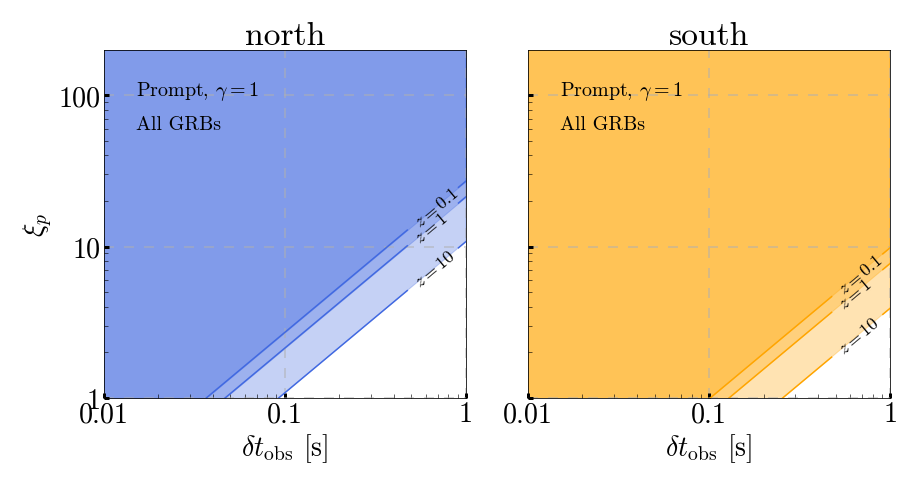}
  \includegraphics[width=1.\linewidth]{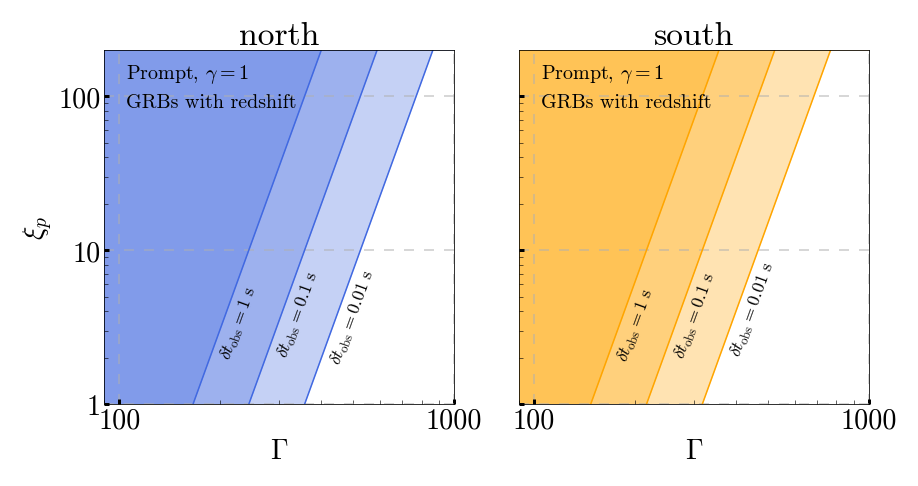}
  \caption{$90\%$ CL upper limits for the baryon loading factor $\xi_p$ in the Northern (left) and Southern (right) Sky. These constraints are obtained with the stacking analysis of the prompt catalog (upper plots) and subcatalog of GRBs with available redshift (bottom plots), assuming a neutrino spectral index $\gamma=1$. Excluded values are shown as shaded regions for different hypothesis of the redshift $z$ (upper plots) and of the timescale variability $\delta t_\mathrm{obs}$ (bottom plots).}
  \label{fig:gamma1_prompt_constraint}
\end{figure}

\begin{figure}[!htb]
  \centering
  \includegraphics[width=1.\linewidth]{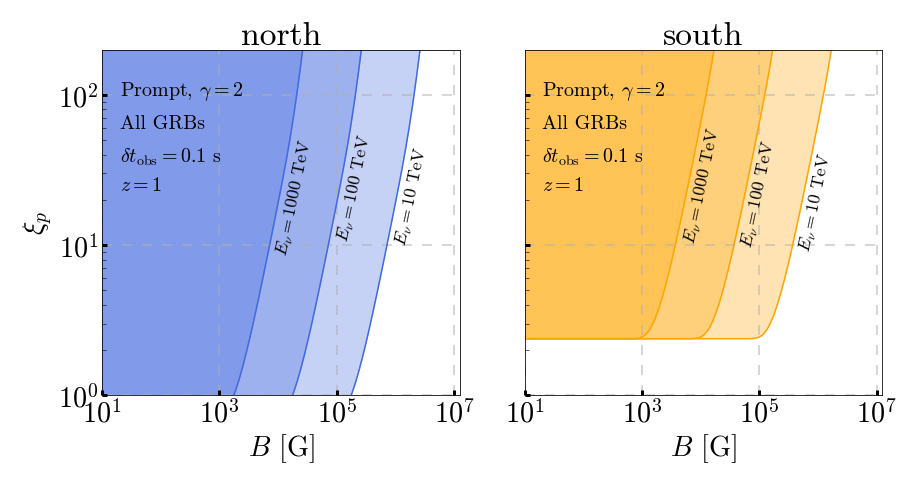}
  \caption{$90\%$ CL upper limits for the baryon loading factor $\xi_p$ in the Northern (left) and Southern (right) Sky as a function of the magnetic field $B$. These constraints are obtained with the stacking analysis of the prompt catalog, assuming a neutrino spectral index $\gamma=2$. Excluded values are shown as shaded regions for different values of the neutrino energy $E_\nu$. Typical values of the timescale variability $\delta t_\mathrm{obs}=0.1$~s and redshift $z=1$ are used to compute these limits. It is worth notice that the asymptotic value of $\xi_p$ in the low-magnetic field regime in the Southern Sky is consistent with the limits computed in Fig.~\ref{fig:gamma2_prompt_constraint} in the same hemisphere, assuming the same parameters.}
  \label{fig:gamma2_Bfield_constraint}
\end{figure}

We obtain constraints on the baryon loading factor $\xi_p$ of the prompt emission as a function of the timescale variability $\delta t_\mathrm{obs}$ and bulk Lorentz factor $\Gamma$. Such constraints are shown in Fig.~\ref{fig:gamma2_prompt_constraint} (assuming a neutrino spectral index $\gamma=2$) and \ref{fig:gamma1_prompt_constraint} (assuming a neutrino spectral index $\gamma=1$), for the two hemispheres and considering measured redshifts or benchmark values (when not measured), as well as some values of the variability timescale $\delta t_\mathrm{obs}$.
Our analysis generally limits the baryon loading factor to $\xi \lesssim 10$, unless extreme GRB values are considered (e.g. $\Gamma>500$, $\delta t_\mathrm{obs}\sim 1$~s, $z\sim10$).
This seems to disfavor a baryonic origin of the GRB prompt, as also indicated by the absence of significant astrophysical neutrino results. Furthermore, if the synchrotron cooling suppression factors in Eq.~\ref{eq:f_sync} are not neglected, we can use Eq.~\ref{eq:nuFluence} to similarly constrain the baryon loading factor $\xi_p$ as a function of the magnetic field $B$, assuming typical values of $\delta t_\mathrm{obs}=0.1$~s and $z=1$. This is shown in Fig.~\ref{fig:gamma2_Bfield_constraint}, and indicates that for neutrino energies $\lesssim100$~TeV (where most of the IceCube events are observed), a large magnetic field $B\gtrsim10^5$~G must be considered to accommodate the observations reported in this work. This might hint at the fact that GRB jets are mostly magnetic-dominated, as already proposed by some authors~\citep{Usov:1992zd,Thompson:1994zh,Lyutikov_2003astro.ph.12347L,Ghisellini2020}.

\begin{figure}[!htb]
  \centering
  \includegraphics[width=1.\linewidth]{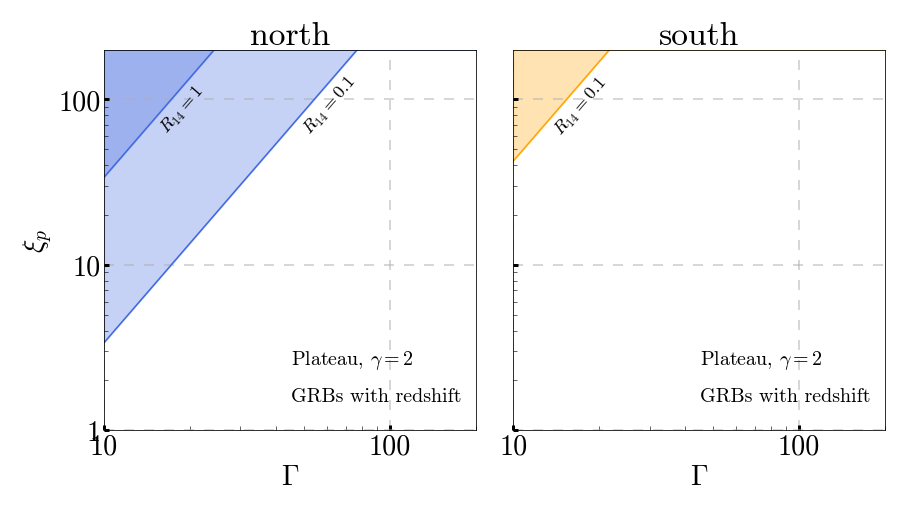}
  \includegraphics[width=1.\linewidth]{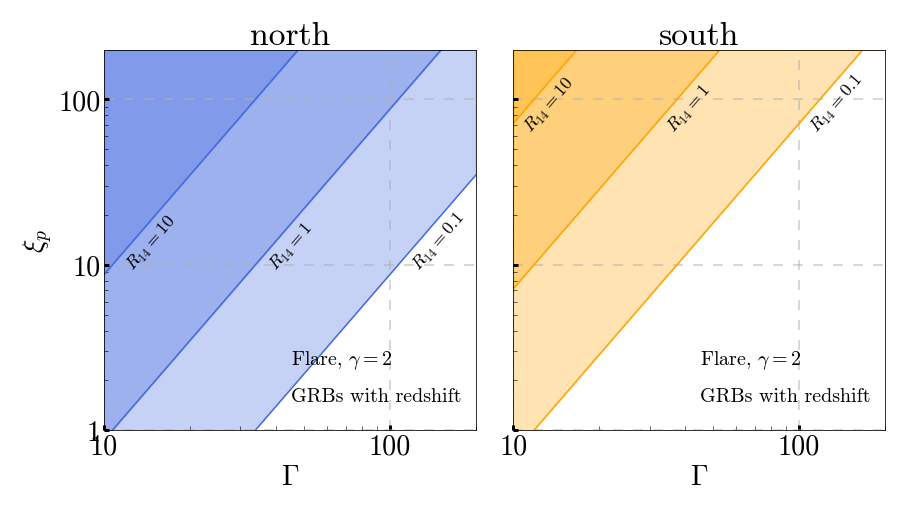}
  \caption{$90\%$ CL upper limits for the baryon loading factor $\xi_p$ in the Northern (left) and Southern (right) Sky. These constraints are obtained with the stacking analysis of the plateau (upper plots) and flare (bottom plots) subcatalogs of GRBs with available redshift, assuming $\gamma=2$. Excluded values are shown as shaded regions for different hypotheses on the normalized radius of the GRB emission site.}
  \label{fig:gamma2_Xray_constraint}
\end{figure}

\begin{figure}[!htb]
  \centering
  \includegraphics[width=1.\linewidth]{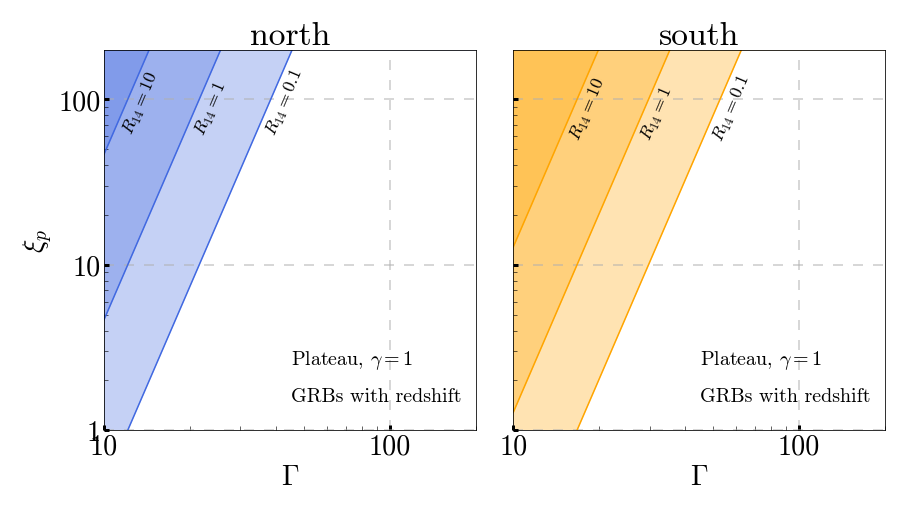}
  \includegraphics[width=1.\linewidth]{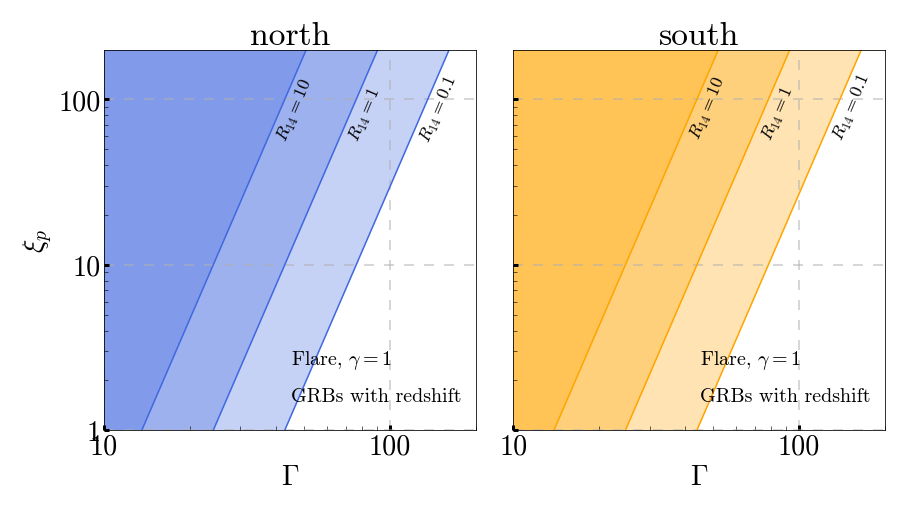}
  \caption{$90\%$ CL upper limits for the baryon loading factor $\xi_p$ in the Northern (left) and Southern (right) Sky. These constraints are obtained with the stacking analysis of the plateau (upper plots) and flare (bottom plots) subcatalogs of GRBs with available redshift, assuming $\gamma=1$. Excluded values are shown as shaded regions for different hypotheses on the normalized radius of the GRB emission site $R_{14}$ (see Sec.~\ref{sec:model}).}
  \label{fig:gamma1_Xray_constraint}
\end{figure}

We additionally constrain the baryon loading factor of the plateau and flare subcatalogs (with available redshift) as a function of the Lorentz factor $\Gamma$, for some values of the normalized radius of the GRB emission site $R_{14}$. In this case, we cannot use the empirical correlations observed for the prompt phase and mentioned above, but we exploit all the available observations of the X-ray afterglows, namely the observed luminosity, fluence, and redshift. The constraints of our analyses are shown in Fig.~\ref{fig:gamma2_Xray_constraint} (assuming a neutrino spectral index $\gamma=2$) and \ref{fig:gamma1_Xray_constraint} (assuming a neutrino spectral index $\gamma=1$).
It should be noticed that in the case of X-ray afterglows, the variability range of the bulk Lorentz factor is typically lower ($10<\Gamma<10^2$) than for prompt ($10^2<\Gamma<10^3$)~\citep{Kimura_nu_from_GRB:2022zyg}. 
The constraints on the X-ray afterglows are looser than those on the prompt phase, and the analysis does not exclude a possible baryonic origin of the plateau and flare emission. 

The non-observation of an astrophysical neutrino signal from GRB prompt and afterglow phases is compatible with previous IceCube results, and with the estimated upper limits of GRB neutrinos to the diffuse flux of $\lesssim 1\%$~\citep{Aartsen_2017,GRB_IC_2022}.
    \section{Conclusions}
\label{sec:conclusions}

In this paper, we analyzed the 10-year IceCube data publicly available~\citep{IceCube_data_release:2021xar}. We used the PSLab code~\citep{PSLab}, recently released by the IceCube collaboration, to perform an unbinned maximum-likelihood search for spatial and temporal coincidence of astrophysical neutrinos from individual GRBs with X-ray flare and X-ray plateau afterglows. We also performed a stacking search for a cumulative neutrino excess from a flare, a plateau, and a prompt catalog. Unlike previous IceCube stacking searches, that assumed the same fluence at Earth for all GRBs, in our analysis we proposed a stacking scheme based on physically motivated GRB weights and observational properties. Furthermore, while past searches adopted the same benchmark values of relevant parameters for all the GRBs, we fit and used empirical relationships to estimate the GRB luminosities and Lorentz factors in the analysis of the prompt catalog when the redshift is not available.
Since such parameters can vary across several orders of magnitude, our novel approach improves the physical reliability of the results.

We did not find any statistically significant neutrino correlations in our searches, consistently with previous IceCube results~\citep{Aartsen_2017,GRB_IC_2022}. The non-observation of neutrinos is therefore used to place upper limits on the individual and stacking flux, and to constrain general parameters of fundamental interest, such as the baryon loading, the Lorentz factor, and the magnetic field of the jet in a general single-zone fireball model. The results of the stacking analysis for the prompt phase suggests a typical baryon loading factor $\xi_p\lesssim10$, thus excluding the hypothesis of a baryonic origin of the jets and supporting the scenario of a magnetic-dominated ejecta. In this scenario, GeV neutrinos might be expected as a consequence of the highly efficient synchrotron cooling of pions and muons, and potentially detectable by IceCube DeepCore and KM3NeT/ORCA~\citep{GeV_nu_PhysRevD.105.083023}. The constraints of the afterglow phase are less tight, but it is worth mentioning that this is the first analysis that specifically targets GRB X-ray plateaus and flares.


    
    \begin{acknowledgements}
        We thank the IceCube Collaboration which offered the optimal environment for the analysis code development and for its many applications and for having released the 10-year data sample. This work also made use of data supplied by the UK Swift Science Data Centre at the University of Leicester. We are thankful to Christopher Wiebusch for his detailed comments to the paper, and to the Publication Committee of IceCube for having discussed it. GO thanks Annalisa Celotti and Dafne Guetta for fruitful discussions. The work was financed by the Swiss National Foundation grant n. 200020\_178918 and by the University of Geneva. GO and MB acknowledge financial support from the AHEAD2020 project (grant agreement n. 871158). BB and MB acknowledge financial support from MUR (PRIN 2017325 grant 20179ZF5KS).
    \end{acknowledgements}
    
    \clearpage
    \bibliographystyle{aa}
    \bibliography{references}

\end{document}